%% file: main.tex
\def\editmode{0}
\def\reportmode{1}

\def\bibfilenames{wisecart_bibliography}

\if\reportmode1
\documentclass[12pt,onecolumn,draftclsnofoot]{IEEEtran}
\else
\documentclass[journal]{IEEEtran}
\fi
\usepackage{amsthm} 
\newtheorem{theorem}{Theorem}
\newtheorem{lemma}{Lemma}
\theoremstyle{definition}
\newtheorem{definition}{Definition}
\newcommand{\R}{\mathbb{R}}
\if\editmode1  
\usepackage[normalem]{ulem}
\usepackage[backend=bibtex,style=alphabetic,sorting=debug,maxbibnames=99]{biblatex}
\DeclareFieldFormat{labelalpha}{\thefield{entrykey}}
\DeclareFieldFormat{extraalpha}{}
\bibliography{\bibfilenames}
\newcommand{\cmt}[1]{\noindent\textcolor{lightgreen}{\uline{[#1]}}} 
\newcommand{\acom}[1]{\textcolor{red}{\textbf{[#1]}}}
\newenvironment{myitemize}{\begin{itemize}}{\end{itemize}}
\newcommand{\myitem}{\item}

\else
\usepackage{cite}
\newcommand{\cmt}[1]{} 
\newcommand{\acom}[1]{} 
\newenvironment{myitemize}{}{}
\newcommand{\myitem}{}

\newcommand{\myacom}[1]{\textcolor{red}{\textbf{[#1]}}}
\fi 

\usepackage[acronym]{glossaries}
\newacronym{DtD}{D2D}{device-to-device}
\newacronym{CSI}{CSI}{channel state information}
\newacronym{BS}{BS}{base station}
\newacronym{CU}{CU}{cellular user}
\newacronym{UE}{UE}{user equipment}
\newacronym{CUs}{CUs}{cellular users}
\newacronym{QoS}{QoS}{quality of service}
\newacronym{SINR}{SINR}{signal to interference plus noise ratio}
\newacronym{LCC}{ICSI}{imperfect CSI}
\newacronym{FCC}{PCSI}{perfect CSI}
\newacronym{ERM}{ERM}{expected rate maximization}
\newacronym{MRM}{MRM}{minimum guaranteed rate maximization}

\usepackage{diagbox}
\usepackage{algorithm}
\usepackage{algpseudocode}

\usepackage{tikz}
\usetikzlibrary{patterns}
\usetikzlibrary{arrows.meta}
\usepackage{pgfplots}
\pgfplotsset{compat=1.11}
\usepgfplotslibrary{fillbetween}
\usetikzlibrary{intersections}
\usepackage{amsthm}
\usepackage{graphicx}
\usepackage{epstopdf}

\input{include}

\begin{document}
\title{Robust Underlay Device-to-Device Communications on Multiple Channels}
\author{Mohamed Elnourani,~\IEEEmembership{Student Member,~IEEE,} Siddharth Deshmukh,~\IEEEmembership{Member,~IEEE,} Baltasar Beferull-Lozano,~\IEEEmembership{Senior Member,~IEEE,} Daniel Romero,~\IEEEmembership{Member,~IEEE}\acom{\today} \thanks{This work was supported by the FRIPRO TOPPFORSK grant WISECART 250910/F20 from the Research Council of Norway.}
}
\maketitle
\vspace{-10mm}
\begin{abstract}
Most recent works in \gls{DtD} underlay communications focus on the optimization
of either power or channel allocation to improve the spectral efficiency, and typically consider
uplink and downlink separately. Further, several of them also assume perfect knowledge of channel-state-information (CSI). In this paper, we formulate a joint uplink and downlink resource allocation scheme, which assigns both power and channel resources to D2D pairs and cellular users in an underlay network scenario.
The objective is to maximize the overall network rate while maintaining fairness among the D2D pairs.
In addition, we also consider imperfect CSI, where we guarantee a certain outage probability to maintain the desired quality-of-service (QoS). The resulting problem is a mixed integer non-convex optimization problem and we propose both centralized and decentralized algorithms to solve it, using convex relaxation, fractional programming, and alternating optimization. In the decentralized setting, the computational load is distributed among the D2D pairs and the base station, keeping also a low communication overhead. Moreover, we also provide a theoretical convergence analysis, including also the rate of convergence to stationary points. The proposed algorithms have been experimentally tested in a simulation environment, showing their favorable performance, as compared with the state-of-the-art alternatives.

\end{abstract}




\section{Introduction}
\cmt{Motivation/overview} 
\cmt{(from general to specific, ending with the goal of the paper)}
\begin{myitemize}
\myitem\cmt{increase spatial efficiency}Throughput demands of cellular communications have been exponentially increasing over the last few  years ~\cite{obile2016ericsson,networking2017ciscoglobal}. Classical techniques to enhance the \emph{spectral} efficiency of point-to-point links can not satisfy this demand, since existing systems already achieve rates close to the single-user channel capacity~\cite{csiszar2011information}. For this reason, recent research efforts target to improve the \emph{spatial} efficiency.  \myitem\cmt{D2D}\gls{DtD} communications constitute a prominent example, where cellular devices are allowed to communicate directly
with each other without passing their messages through
the \Gls{BS}~\cite{kaufman2008cellular,asadi2014survey,liu2015device,jameel2018survey}. This
paradigm entails higher throughput and  lower latency in the communication for two
reasons:
\myitem\cmt{benefit 1:general}first, a traditional cellular
communication between two devices requires one time slot in the uplink
and one time slot in the downlink, whereas a single time slot suffices
in \gls{DtD} communications.
\myitem\cmt{benefit 2:underlay}Second, the time slot used by a traditional \gls{CU} can be simultaneously used by a D2D pair in a sufficiently distant part
of the cell, a technique termed \emph{underlay}.
\myitem\cmt{power/channel allocation in underlay D2D}
In order to fully exploit the potential of underlay \gls{DtD} communications,
algorithms that provide a judicious assignment of cellular
sub-channels (e.g. resource blocks in LTE or time slots at each
frequency in general) to \gls{DtD} users as well as a prudent power
control mechanism that precludes detrimental interference to \gls{CUs} are necessary. These research challenges constitute the main motivation of this paper.
\end{myitemize}
\cmt{Literature}
\begin{myitemize}
\myitem\cmt{\cite{kaufman2008cellular}}

\myitem\cmt{random channel allocation}Early works on  D2D
communications rely on simplistic channel assignment schemes, where
each D2D pair communicates through a randomly selected cellular sub-channel
(hereafter referred to as \emph{channel} for simplicity). This is the case in
\begin{myitemize}\myitem\cmt{\cite{doppler2010mode}}\cite{doppler2010mode}, where the effects of selecting
a channel with poor quality are addressed by choosing the best among the following operating modes: underlay mode; overlay mode (the D2D pair is assigned a channel that is unused by the CUs); and cellular mode (the D2D pair operates as a regular cellular user).
\myitem \cmt{\cite{sakr2015cognitive}}
Alternatively, a scheme is proposed in \cite{sakr2015cognitive} which allows D2D pairs to perform spectrum sensing and opportunistically communicate over a single channel randomly selected by the BS for all the D2D communications.
\myitem\cmt{summary of limitations} These research works suffer from two limitations:
\begin{myitemize}
\myitem\cmt{channel assignment}(i) random allocation of channels result in a sub-optimal throughput which could be improved by leveraging different degrees of channel-state information;
\myitem\cmt{power control}(ii) they do not provide any mechanism to adjust the
transmit power of D2D terminals, which generally results in a reduced throughput due to increase in interference.
\end{myitemize}
\end{myitemize}

\myitem\cmt{channel allocation}
\begin{myitemize}
\myitem \cmt{\cite{xu2012interference,xu2012resource}}
Few works also consider performing channel assignment to the D2D pairs for underlay communication.  In~\cite{xu2012interference,xu2012resource}, instead of randomly assigning channels, channels are assigned to the D2D pairs using auction games while maintaining the fairness in the number of channels assigned to each D2D pair. Similarly, \cite{chen2018resource} proposes channel assignment to D2D pairs utilizing a coalition-forming game model. Here, millimeter-wave spectrum is also considered as an overlay option for \gls{DtD} pairs. 
However, it can be noted that these schemes only perform channel assignment and avoid controlling the transmit power, which limits the achievable throughput of the overall network.
\end{myitemize}

\myitem\cmt{power allocation}
\begin{myitemize}
\myitem \cmt{\cite{yin2013distributed,hajiaghajani2016joint}}In order to circumvent above limitations, a Stackelberg game based approach is proposed in \cite{yin2013distributed} where each
D2D pair simultaneously transmits in all cellular channels and compete non-cooperatively to adjust the transmit power. Here, the BS penalizes the \gls{DtD} pairs if they generate harmful interference to the cellular communication. 
\myitem \cmt{\cite{alihemmati2017power}}The optimization of the transmit power while ensuring minimum \gls{SINR} requirements is also investigated in \cite{alihemmati2017power}, however, as long as the \gls{SINR} requirements are satisfied, \gls{DtD} pairs are allowed to transmit in all channels.
\myitem \cmt{\cite{abrardo2017distributed}} In an alternative approach, distributed optimization for power allocation is investigated  in \cite{abrardo2017distributed} for both overlay and underlay scenarios. 
To summarize, all the works mentioned so far perform either channel assignment or power allocation, but not both.

\end{myitemize}

\myitem\cmt{joint channel and power alloc.}
Few research works consider jointly optimizing channel assignment and power allocation, as they seem to show strong dependency.
\begin{myitemize}\myitem \cmt{\cite{feng2013device}}This joint optimization is considered in \cite{feng2013device,yu2014joint,jiang2016energy,dominic2016distributed} 
\myitem\cmt{summary:limitations}
but they restrict D2D users to access at most one cellular channel. 
However, the work in \cite{hajiaghajani2016joint,yuan2018iterative,guo2018fairness} allow assignment of multiple channels to each D2D pair. 
Notice that these schemes propose to use either uplink or downlink spectrum for D2D communications. \end{myitemize} Some recent research works also consider both uplink and downlink spectrum for allocating resources to D2D pairs.
\myitem \cmt{\cite{zhao2017gain,kai2018joint,kai2019joint}}In \cite{zhao2017gain,kai2018joint,kai2019joint}, both uplink and downlink spectra are considered in their formulation; however, they limit the assignment to at most one channel to each D2D pair.

Another important point to note is that all of the previously mentioned works assume the availability of perfect \gls{CSI}. From a practical prospective, obtaining perfect \gls{CSI} for D2D communications requires a lot of cooperation between all \gls{DtD} pairs and \gls{CUs}; thus adds a substantial amount of communication overhead. Some other recent works have also investigated problems that guarantee certain QoS parameters under the scenario of imperfect CSI for underlay \gls{DtD} communications. In \myitem \cmt{\cite{feng2016qos,feng2013optimal,thieu2018outage}} \cite{feng2013optimal,feng2016qos,thieu2018outage}, power allocation and channel assignment are considered under imperfect \gls{CSI}. However, the analysis, once again, restricts D2D pairs to access at most one cellular channel. Table \ref{tab:Refs} list some of the presented works that jointly perform channel assignment and power allocation compared with our proposed scheme.
{\small
\setlength{\textfloatsep}{0pt}
\begin{table}[t]
    \centering
    \begin{tabular}{|c||c|c|c|} \hline
         Works & Multiple channels & Joint UL and DL & CSI uncertainty \\ \hline
         \cite{feng2013device,yu2014joint,jiang2016energy,dominic2016distributed} &  & &  \\\hline
         \cite{hajiaghajani2016joint,yuan2018iterative,guo2018fairness} & X & & \\\hline
         \cite{zhao2017gain,kai2018joint,kai2019joint} & & X & \\\hline
         \cite{feng2013optimal,feng2016qos,thieu2018outage} &  &  & X \\\hline \hline
         Proposed & X & X & X \\\hline
    \end{tabular}
    \vspace{-4mm}
    \caption{Selected works that jointly perform channel assignment and power allocation}
    \label{tab:Refs}
    \vspace{-8mm}
\end{table}}
\acom{Even though the problem have been extended to the case of D2D pairs having multiple antennas\footnote{This is part of our future work.} \cite{lin2016beamforming,Wei2013beamforming}, our work shows that even in the case of single antenna, the resulting problem contains already the main ingredients in terms of resource allocation and has a sufficiently high complexity to be solved that deserves a thorough analysis and understanding, before considering the multiple antenna case.}

In this paper, we consider the resource allocation involving both uplink and downlink. Fig. \ref{sys_model} illustrates the potential of such an approach. Here for instance, if channel CH$_1$ is assigned to D2D pair $1$, it is better to use downlink spectrum since it observes less cellular interference. Similarly, if channel CH$_2$ is assigned to D2D pair $2$, it is better to use uplink spectrum. Further, one can notice it is better to assign channel CH$_2$ to D2D pair $1$ and channel CH$_1$ to  D2D pair $2$ if only uplink spectrum is available for underlay communication. Furthermore, with the possibility of assigning multiple channels to each D2D pair, the number of potential choices increases substantially, allowing a more favourable channel assignment and power allocation.
\begin{figure}
\centering
\includegraphics[width=0.8\textwidth]{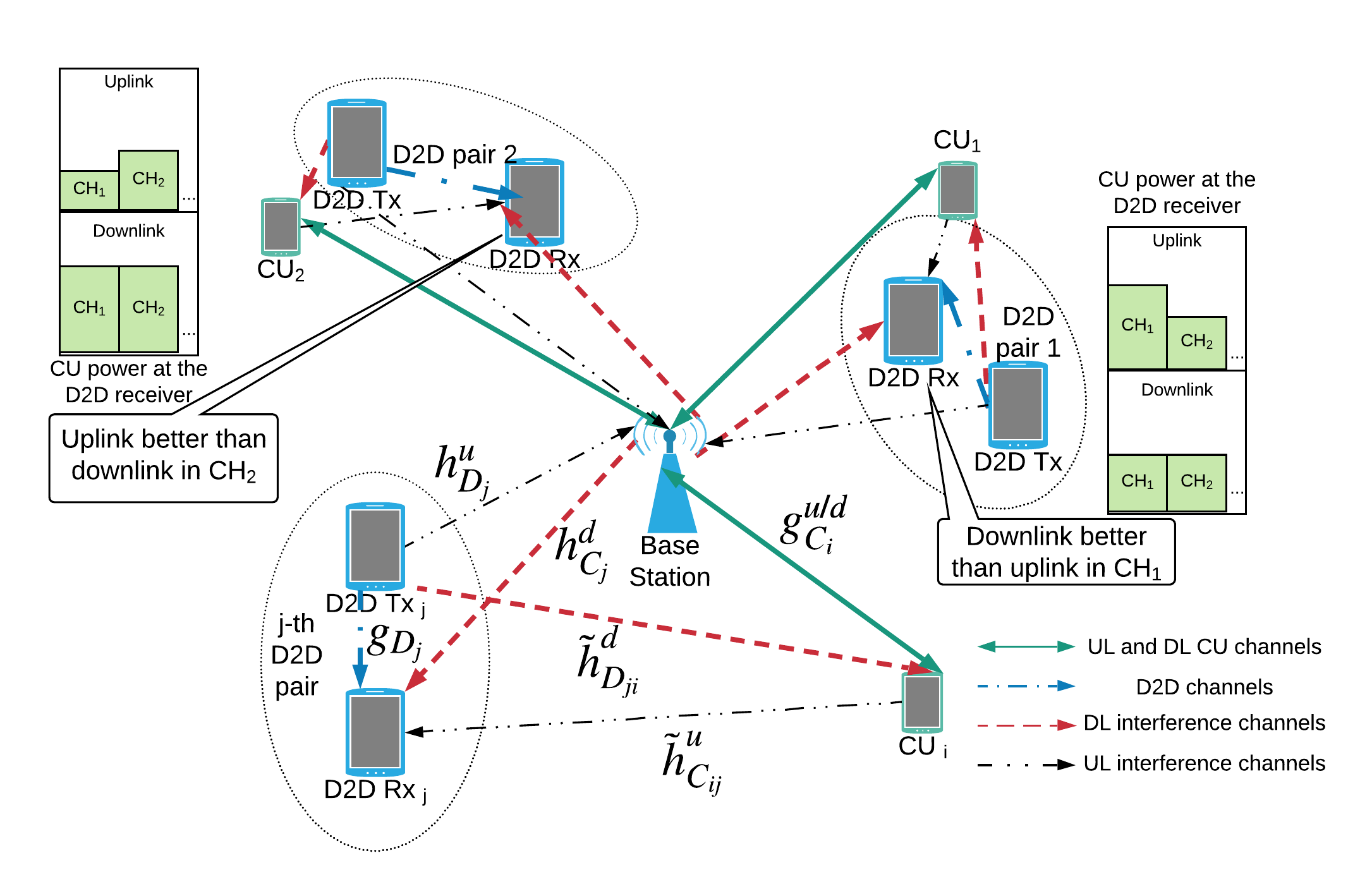}
\caption{Overall Network model}
\label{sys_model}
\vspace{-8mm}
\end{figure}
In conclusion, no existing work
provides a joint channel assignment and power allocation scheme that satisfies all of the following requirements: 
(i) considers both uplink and downlink spectrum;
(ii) accounts for uncertainties in CSI and thus obtains a robust resource allocation solution;
and (iii) D2D pairs can simultaneously operate on more than one cellular channel, which is of special interest in areas of high \gls{CU} density.\end{myitemize}
This paper addresses the above mentioned limitations 
and provides the following research contributions:
\begin{itemize}
\item\cmt{formulation} We propose a joint uplink and downlink resource allocation scheme, which assigns both power and channel resources to D2D pairs and CUs. The objective of this scheme is to maximize the total network rate while maintaining fairness in the channel assignment among the D2D pairs. In addition, the scheme also allows assigning multiple channels to each D2D pair. Moreover, the proposed scheme also accounts for uncertainties in the channels by introducing probabilistic constraints that guarantee the desired outage probabilities.
\item\cmt{Algo.} We propose a computationally efficient solution for the resulting problem, even though it is a mixed integer non-convex optimization problem, which involves exponential complexity to compute the optimal solution. We first show that without loss of optimality, the overall problem can be decomposed into several power allocation subproblems and a channel assignment problem. The solution of the power allocation sub-problems in the case of perfect \gls{CSI} is obtained in closed form, whereas in the scenario of imperfect CSI, a quadratic transformation and alternating optimization methods are proposed. The proposed algorithms can be implemented centrally at the \gls{BS}.
\item We also propose decentralized algorithms that reduce the computational load at the BS by solving each power allocation sub-problem in parallel at the corresponding D2D pair. Moreover, some of the computations for the channel assignment problem are also performed by the D2D pairs. Furthermore, \acom{since all iterates end with feasible solutions,}the communication can start immediately after the first iteration without waiting for the algorithms convergence.
\item\cmt{Conv.} We provide convergence guarantees to stationary points for all of our algorithms and we show linear convergence rate for some of the considered cases and sub-linear convergence rates for other cases.
\item\cmt{simulations} Extensive simulations are also presented to demonstrate the
advantages of the proposed method as compared to current state-of-the-art alternatives.
\end{itemize}
%
\cmt{paper structure}
\acom{The rest of this paper is structured as follows: Section \ref{sec:Sys_Mod} describes the system model followed by analysis of solutions to the separate uplink and downlink resource allocation problems in section \ref{sec:SepDU}.
Joint uplink and downlink resource allocation is described in section \ref{sec:JointUD} and decentralized solutions are discussed in section \ref{sec:Dec}. 
The simulation parameters and results are presented in section \ref{sec:Sim}. Finally, section \ref{sec:Conc} presents the conclusion and some future directions.}
%
%
\section{System Model and Problem Statement}
\label{sec:Sys_Mod}
\cmt{Cellular network}
\begin{myitemize}
\myitem{\cmt{Underlay}}
Consider an underlay \gls{DtD} communication scenario in which both uplink and downlink cellular channels are accessible to \gls{DtD} pairs. 
 In the following description, we describe the considered communication scenario, which is also depicted in Fig. \ref{sys_model}.
\par \textit{Cellular network configuration:}
  \myitem{\cmt{cell}} We consider a cell (or sector) of a cellular network
  in which the serving \gls{BS} and the associated \gls{CUs} communicate via $N_C^{(u)}$ uplink and $N_C^{(d)}$ downlink channels, respectively\footnote{Recall that channel in this context may
  stand for resource blocks, time slots, and so   on.}. 
  Considering the worst case underlay scenario, we assume, without loss of generality, a fully loaded cellular communication scenario in which all uplink and downlink channels are assigned to CUs. For notational convenience, the set of CUs communicating in respective uplink and downlink channels are indexed as $\mathcal{C}^{(u)}$ = $\left\{1, ...,N_C^{(u)}\right\}$ and $\mathcal{C}^{(d)}$ = $\left\{1, ...,N_C^{(d)}\right\}$.
\end{myitemize}

\cmt{D2D devices}
\begin{myitemize}
\myitem\textit{D2D communication configuration:}
  \myitem\cmt{D2D pairs} Next, we assume that $N_D$ \gls{DtD} pairs $\left(\right.$indexed by $j\in \mathcal{D}$ =
  $\left.\left\{1, ...,N_D\right\}\right)$ desire to communicate \myitem\cmt{resource sharing}over the aforementioned downlink and uplink channels in an underlay configuration, i.e., simultaneously on the same uplink and downlink channels assigned to the \gls{CUs}. 
  \myitem\cmt{betas}The assignment of uplink or downlink channels to D2D pairs is
  represented by the indicator variables $\{\beta^{(u)}_{i,j}\}$ and $\{\beta^{(d)}_{i,j}\}$, respectively,  where $j$ denotes D2D pair ($j \in \mathcal{D}$) and $i$ denotes either uplink or downlink channel $\left(i \in \mathcal{C}^{(u)}~ or ~\mathcal{C}^{(d)}\right)$. Here $\beta^{(u)}_{i,j}=1$ or $\beta^{(d)}_{i,j} = 1$ when the $j$-th \gls{DtD} pair accesses the $i$-th uplink or downlink channel.
  In order to improve throughput of the \gls{DtD} pairs, we further assume that each \gls{DtD} pair can access multiple channels at the same time. However, in-order to restrict interference among \gls{DtD} pairs, we assume that each channel can be used by at most one \gls{DtD} pair, this can be expressed as $\sum\limits_{j=1}^{N_D}\beta^{(u)}_{i,j}\leq1,~\sum\limits_{j=1}^{N_D}\beta^{(d)}_{i,j}\leq1~\forall~i$. 
  \myitem In addition, to reduce hardware complexity, we further assume that each \gls{DtD} pair can have access to multiple channels in either downlink or uplink spectrum band \cite{zhao2017gain,kai2018joint}, which can be expressed as $\sum\limits_{i=1}^{N_C^{(u)}}\beta^{(u)}_{i,j}\times \sum\limits_{i=1}^{N_C^{(d)}}\beta^{(d)}_{i,j} = 0,~\forall j$.
\end{myitemize}

\cmt{quantities of interest}
   \begin{myitemize}
\myitem \textit{Communication channels:}  
  \myitem\cmt{channel gains} First, we define channel gains in the uplink access. Let 
  \begin{myitemize}
  \myitem{$g^{(u)}_{C_i}$} denote the channel gain from the $i$-th CU to the BS and \myitem{$h^{(u)}_{C_{i,j}}$} denote the channel gain of the interference link from the $i$-th \gls{CU} to the $j$-th \gls{DtD} pair receiver. Similarly, let \myitem{$g_{D_j}$} denote the channel gain between transmitter and receiver of the $j$-th \gls{DtD} pair and \myitem{$h^{(u)}_{D_j}$} denote the channel gain of the interference link from the transmitter of the $j$-th \gls{DtD} pair to the \gls{BS}. Next, for downlink access, let \myitem{$g^{(d)}_{C_i}$} denote the channel gain from the \gls{BS} to the $i$-th \gls{CU} and \myitem{$h^{(d)}_{C_{j}}$} denote the channel gain of the interference link from the BS to the $j$-th \gls{DtD} pair. Finally, let  \myitem{$h^{(d)}_{D_{j,i}}$} denote the channel gain of the interference link from the transmitter of the $j$-th \gls{DtD} pair to the $i$-th \gls{CU}.  Here we assume that the interference channel gains  affecting the \gls{CUs} are estimated with minimum cooperation from the \gls{CUs}; thus, gains of these interference links are assumed to be modeled as random variables, denoted respectively by $\tilde{h}^{(u)}_{C_{i,j}}$ and $\tilde{h}^{(d)}_{D_{j,i}}$. Finally, additive noise observed in individual channels is assumed to 
  have a known power $N_0$. Note that the noise and all channel gains are assume to be frequency flat to simplify the notations; however, the proposed scheme carries over immediately to the frequency selective scenario.
 \end{myitemize}

\begin{myitemize}
\myitem\textit{Transmit power constraints:}
\myitem\cmt{power} Considering the limited power available at the mobile devices, the transmit power of the $j$-th \gls{DtD} pair when assigned to the $i$-th uplink or downlink channel, denoted as $P^{(u)}_{D_{j,i}}$ and $P^{(d)}_{D_{j,i}}$ is constrained as $0\leq P^{(u)}_{D_{j,i}}\leq P_{D_{max}},~0\leq P^{(d)}_{D_{j,i}}\leq P_{D_{max}}$. Similarly, the transmit power of the \gls{CU} on the $i$-th uplink channel and of the \gls{BS} on the $i$-th downlink channels are constrained, respectively, as $0\leq P^{(u)}_{C_{i}}\leq P^{(u)}_{C_{max}}$ and $0\leq P^{(d)}_{C_{i}}\leq P^{(d)}_{C_{max}}$. Note that $P^{(u)}_{C_{max}},~P^{(d)}_{C_{max}},~$and$~P_{D_{max}}$ are assumed to be the same for all CUs and D2D pairs to simplify the notations, however, once again the proposed scheme carries over immediately to the scenario where they are different.
\end{myitemize}
\begin{myitemize}

\myitem\textit{Achievable rates:}
\myitem\cmt{rates} Here, we first present the achievable rates for \gls{DtD} underlay communication on downlink channels and then extend our discussion for underlay on uplink channels. Let \myitem{$R^{(d)}_{C_{i,j}}$} and \myitem{$R^{(d)}_{D_{j,i}}$} denote the rate of the $i$-th \gls{CU} and of the $j$-th \gls{DtD} pair when sharing the downlink channel, which are respectively given as:
\begin{equation*}
\begin{array}{ll}
R^{(d)}_{C_{i,j}} =\log_2\left(1+\dfrac{P^{(d)}_{C_{i}}g^{(d)}_{C_i}}{N_0+P^{(d)}_{D_{j,i}}\tilde{h}^{(d)}_{D_{j,i}}}\right),~ 
R^{(d)}_{D_{j,i}} =\log_2\left(1+\dfrac{P^{(d)}_{D_{j,i}}g_{D_j}}{N_0+P^{(d)}_{C_{i}}{h}^{(d)}_{C_j}}\right).
\end{array}
\end{equation*}
When the $i$-th \gls{CU} does not share the downlink channel, the achievable rate denoted by \myitem{$R^{(d)}_{C_{i,0}}$} is given as:
\begin{equation*}
R^{(d)}_{C_{i,0}} = \log_2\left(1+\dfrac{P^{(d)}_{C_{\max}}g^{(d)}_{C_i}}{N_0}\right).
\end{equation*}
Thus, the gain in rate when the $i$-th \gls{CU}  shares channel with the $j$-th \gls{DtD} pair can be stated as, $v^{(d)}_{i,j}=R^{(d)}_{C_{i,j}}+R^{(d)}_{D_{j,i}}-R^{(d)}_{C_{i,0}}$. 
Finally, the overall network rate in the downlink can be stated as:
\begin{equation}
\label{eq:totalrateasgainsum}
R^{(d)}(\bm B^{(d)},\bold{p_{C}}^{(d)} , \bm{p_{D}}^{(d)}) = \sum_{i\in\mathcal{C}^{(d)}}\left[\sum_{j\in\mathcal{D}} \beta^{(d)}_{i,j}v^{(d)}_{i,j} + R^{(d)}_{C_{i,0}} \right].
\end{equation}
where $\bm B^{(d)}\triangleq [\beta_{i,j}^{(d)}],~\bold{p_C}^{(d)}\triangleq [P_{C_{1}}^{(d)}, P_{C_{2}}^{(d)},\dots,P_{C_{N_C}}^{(d)}]^T,~\bm{P_D}^{(d)} \triangleq [P_{D{ji}}^{(d)}]$.
\end{myitemize}\\
Similarly, the achievable rates in the uplink channels when sharing the $i$-th CU uplink channel with the $j$-th \gls{DtD} pair can be expressed as:
\begin{equation*}
\begin{array}{ll}
R^{(u)}_{C_{i,j}} =\log_2\left(1+\dfrac{P^{(u)}_{C_{i}}g^{(u)}_{C_i}}{N_0+P^{(u)}_{D_{j,i}}{h}^{(u)}_{D_{j}}}\right),~ 
R^{(u)}_{D_{j,i}} =\log_2\left(1+\dfrac{P^{(u)}_{D_{j,i}}g_{D_j}}{N_0+P^{(u)}_{C_{i}}\tilde{h}^{(u)}_{C_{i,j}}}\right).
\end{array}
\end{equation*}
The achievable rates in the uplink channels without sharing the $i$-th CU uplink channel, the rate gain, as well as the  total rate due to underlay uplink communications can be easily expressed by replacing the superscripts $(d)$ by $(u)$ in the above equations.

\textit{Quality of Service (QoS) requirements:}
\myitem\cmt{Min. SINR} In order to have a successful communication at a receiver node, a minimum signal to interference plus noise (SINR) ratio requirement is imposed in the problem formulation. Thus, for the $i$-th \gls{CU} in the uplink/downlink sharing channel with the $j$-th \gls{DtD} pair, the instantaneous SINR $\eta_{C_{i,j}}^{(u)} \geq \eta^{(u)}_{C_{\min}}$  and $\eta_{C_{i,j}}^{(d)} \geq \eta^{(d)}_{C_{\min}}$, where $\eta^{(u)}_{C_{\min}}$ and $\eta^{(d)}_{C_{\min}}$ are the minimum desired SINR for the \gls{CU} in uplink and donwlink, respectively. Similarly, for the $j$-th \gls{DtD} pair, the instantaneous SINR $\eta_{D_i,j}^{(u)} \geq \eta_{D_{\min}}$, where $\eta_{D_{\min}}$ is the minimum desired SINR for \gls{DtD} pairs in both uplink and downlink. Note thatin order to simplify the notation $\eta^{(u)}_{C_{\min}},~\eta^{(d)}_{C_{\min}},~$and$~\eta_{D_{\min}}$ are also assumed to be the same for all CUs and D2D pairs; however, the scheme carries over immediately to the scenario where they are different. \newline
Since the computations of the SINR for the $j$-th \gls{DtD} pair sharing channel with the $i$-th uplink \gls{CU} involve the random interference channel gain $\tilde{h}^{(u)}_{C{i,j}}$, the minimum SINR requirement can be expressed in terms of a probabilistic constraint as follows: 
\begin{equation*}
Pr\{\eta_{D_{i,j}}^{(u)} \geq \eta^{(u)}_{D_{min}}\} \geq 1-\epsilon ~~~~\forall i \in \mathcal{C}^{(u)}, ~~ \forall j \in \mathcal{D},
\end{equation*}
where $\epsilon$ is the maximum allowed outage probability. Similarly, the minimum SINR requirement for the $i$-th downlink \gls{CU} sharing channel with the $j$-th \gls{DtD} pair can be expressed in terms of a probabilistic constraint as follows:
\begin{equation*}
Pr\{\eta_{C_{i,j}}^{(d)} \geq \eta^{(d)}_{C_{min}}\} \geq 1-\epsilon ~~~~\forall i \in \mathcal{C}^{(d)}, ~~\forall j \in \mathcal{D}.
\end{equation*}
\myitem\cmt{fairness}
\begin{myitemize}
\textit{Fairness in channel assignment to \gls{DtD} pairs:}
\myitem\cmt{definitions}
Let \myitem{$m_{j}$ denotes the number of channels assigned to the $j$-th \gls{DtD} pair.}
\begin{equation*}
m_{j}=\sum\limits_{i_u=1}^{N^{(u)}_C} \sum\limits_{i_d=1}^{N^{(d)}_C}\left(\beta^{(u)}_{i_u,j} + \beta^{(d)}_{i_d,j} \right).
\end{equation*}
Then inspired by the fairness definition in \cite{xu2012interference}, the fairness of a channel allocation can be expressed in terms of a normalized variance from a specified reference assignment \myitem{$m_{0}$} as follows:
\begin{align}
\delta &= \dfrac{{\dfrac{1}{N_D}\sum\limits_{j=1}^{N_D}(m_j-m_0)^2}}{m_0^2} 
\label{eqn:fairness}
\end{align}
\end{myitemize}
\end{myitemize}
\\\textit{Problem statement:}
\cmt{Goal}Given all $g^{(u)}_{C_i},~g^{(d)}_{C_i},~g_{D_j},~h^{(u)}_{D_j},~h^{(d)}_{C_j},~\text{the statistical distribution of }\tilde{h}^{(u)}_{C_{i,j}}$, and $\tilde{h}^{(d)}_{D_{j,i}}
      ~\forall i,j$, as well as $N_0$, $ \eta^C_{\min}$,
      $ \eta^D_{\min}$, $P_{C_{\max}}$, and $P_{D_{\max}}$, the goal
      is to choose $\beta_{i,j}^{(d)},\beta_{i,j}^{(u)},~P_{D_{ji}}^{(d)},~P_{D_{ji}}^{(u)}$, $~P_{C_{i}}^{(d)},~P_{C_{i}}^{(u)}$ to
      maximize the overall rate of the D2D pairs and CUs while
      ensuring fairness among the multiple D2D pairs and preventing
      detrimental interference to CUs.
      
We consider two different scenarios of \gls{DtD} pairs communicating over underlay downlink/ uplink channels: ($\mathbf{S1}$) \gls{DtD} pairs are pre-organized into uplink and downlink groups based on hardware limitations to communicate in either uplink or downlink channels; ($\mathbf{S2}$) The assignment of the \gls{DtD} pairs to either uplink or downlink channels is also part of the optimization problem. Furthermore, in order to reduce the computation load on the \gls{BS}, we also propose decentralized solution for both the scenarios. In both the scenarios, we analyze both perfect and imperfect CSI cases.
\vspace{-4mm}
\section{Separate Downlink and Uplink Resource Allocation}
\label{sec:SepDU}
Recall from Sec. \ref{sec:Sys_Mod} that each D2D pair is allowed to operate either in the uplink or in the downlink, but not in both simultaneously. For the sake of the exposition, this section assumes that the assignment of D2D pairs to either the uplink or downlink is given. Sec. \ref{sec:JointUD} will extend the approach presented in this section to the scenario where such an assignment is not given and therefore becomes part of the resource allocation task.
Thus, there are two pre-organized sets of \gls{DtD} pairs, namely $\mathcal{D}^{(d)}$ and $\mathcal{D}^{(u)}$, intending to communicate in downlink and uplink channels, respectively. Since  the \gls{DtD} pairs are already pre-organized in two separate sets, the joint resource allocation problems simplifies to solving two separate but similar problems: (i) allocating downlink resources to the \gls{DtD} pairs in the set $\mathcal{D}^{(d)}$; (ii) and allocating uplink resources to the \gls{DtD} pairs in the set $\mathcal{D}^{(u)}$. Thus, due to the similarity of the two problems, we only discuss the donwlink resource allocation in this section. Since it introduces no ambiguity and simplifies the notation, in this section, we will drop the superscript denoting uplink and downlink.
\vspace{-4mm}
\subsection{Resource Allocation Under Perfect \gls{CSI} (PCSI)}
\label{subsec:preD}
\acom{details\\}
\cmt{overall optimization problem\\}
Here, we first analyze the ideal scenario in which perfect \gls{CSI} can be exploited to maximize the aggregate throughput of both the \gls{DtD} pairs and \gls{CUs} while ensuring fairness among D2D pairs. To this end, our problem formulation is as follows:
\begin{subequations}
\label{eq:prob}
\begin{align}
\label{eq:probobj}
\maximize_{\bm B,\bold{p_{C}}, \bm{P_{D}}}\quad &
R(\bm B,\bold{p_{C}} , \bm{P_{D}})
-\gamma\delta(\bm B)\\
\text{subject
to}\quad& \beta_{i,j}\in \{0,1\},~\forall
~i,j,~\sum\limits_{j=1}^{N_D}\beta_{i,j}\leq1 ~\forall
~i,,\\
\label{eq:constrpower}
& 0\leq P_{C_{i}}\leq P_{C_{max}},~\forall~i,
~0\leq P_{D_{ji}}\leq P_{D_{max}},~\forall~i,j,\\
\label{eq:constrsnrc}
&~\dfrac{P_{C_{i}}g_{C_i}}{N_0+P_{D_{ji}}h_{D_{j,i}}}\geq \eta^C_{min} \text{
if }\beta_{ij}=1,~\forall ~i,j,\\
\label{eq:constrsnrd}
&\dfrac{P_{D_{ji}}g_{Dj}}{N_0+P_{C_{i}}h_{C_j}}\geq \eta^D_{min} \text{
if }\beta_{ij}=1,~\forall~i,j,
\end{align}
\end{subequations}
where 
the total rate $R(\bm B,\bm{p_{C}},\bm{P_{D}})$ is given by \eqref{eq:totalrateasgainsum}. The fairest resource assignment in this framework corresponds to uniformly distributing the $N_C$ available channels equally over the D2D pairs ($m_0 := {N_C}/{N_D}$). Substituting $m_0$ in \eqref{eqn:fairness}, the fairness in channel allocation $\delta(\bm B)$ can be expressed as:
\begin{align}
\delta(\bm B) &= \dfrac{N_D}{N_C^2}{\sum\limits_{j=1}^{N_D}\left(\sum\limits_{i=1}^{N_C}\beta_{i,j}-N_C/N_D\right)^2}
\end{align}
We consider a user-selected regularization parameter $\gamma>0$ in \eqref{eq:probobj} to balance the rate-fairness trade-off. In general, the highest rate is achieved when all channels are assigned only to D2D pairs with good communications conditions. The fairness in the assignment needs to be enforced by adding a term in the objective function that penalizes unfair assignments.\par
The optimization problem in \eqref{eq:prob} is a non-convex mixed-integer problem and obtaining the optimal solution of such a combinatorial problem will incur an exponential complexity. Next, we show that problem  \eqref{eq:prob} can be decomposed into two steps without loosing optimality: (S1) power allocation; and (S2) channel assignment.  
\par
\begin{myitemize}%
\myitem\cmt{replicating vars}First, consider solving \eqref{eq:probobj} w.r.t. $\bold p_{C}$ for fixed $\bm P_{D}$ and $\bm B$. It can be seen from \eqref{eq:totalrateasgainsum}
that the objective of \eqref{eq:prob} can be written as
$\sum_{i\in\mathcal{C}}\sum_{j\in\mathcal{D}} \beta_{i,j}$ $v_{i,j}(P_{C_{i}},P_{D_{ji}})$
plus some terms that do not depend on $[P_{C_i}]$. Notice that an
equivalent problem can be obtained by replacing $P_{C_i}$ with an artificial auxiliary variable $P_{C_{i,j}}$ in each term $\beta_{i,j}v_{i,j}(P_{C_{i}},P_{D_{ji}})$  and further enforcing the constraint $P_{C_{i,1}}=P_{C_{i,2}}=\ldots=P_{C_{i,N_D}}$ for each
$i$. Then, the modified objective can be expressed as $\sum_{i\in\mathcal{C}}\sum_{j\in\mathcal{D}} \beta_{i,j}v_{i,j}(P_{C_{i,j}},P_{D_{ji}})$ plus terms that do not depend on $[P_{C_{i,j}}]$. Similarly, we can replace $P_{C_i}$ with $P_{C_{i,j}}$ in \eqref{eq:constrsnrc}-\eqref{eq:constrsnrd} and also in \eqref{eq:constrpower} with $ 0\leq P_{C_{i,j}}\leq
P_{C_{max}} \,\forall i,j$ and the resulting problem will be
equivalent to \eqref{eq:prob}.
\myitem\cmt{dropping eq. constraints}%
Thus, except for the recently introduced equality constraints, the objective
and the constraints will only depend on at most one of the
$P_{C_{i,j}}$ for each $i$, specifically the one with $\beta_{i,j}=1$. Hence, 
the equality constraint
$P_{C_{i,1}}=\ldots=P_{C_{i,N_D}}$ can be dropped without
 loss of optimality. 
 \myitem\cmt{recover solution to \eqref{eq:prob}}To recover the optimal $[P^*_{C_i}]$
in \eqref{eq:prob} from the optimal $[P^*_{C_{i,j}}]$
, one just needs to find, for each $i$, the value
of $j$ such that $\beta_{i,j}=1$ and set $P_{C_i}^*=P_{C_{i,j}}^*$. If no
such $j$ exists, i.e.  $\beta_{i,j}=0~~\forall j$, then channel $i$
is not assigned to any D2D pair and the BS can 
transmit with maximum power~$P_{C_i}=P_{C_{\max}}$.
 \myitem\cmt{simplifying \eqref{eq:constrsnrc}-\eqref{eq:constrsnrd}}%
Similarly, without loss of optimality, we can also remove the condition ``if $\beta_{i,j}=1$'' from
\eqref{eq:constrsnrc}-\eqref{eq:constrsnrd}. 
\myitem\cmt{resulting problem}Thus, the resulting problem can be expressed as:
\begin{subequations}
\label{eq:probeq}
\begin{align}
\maximize_{\bm
B,\bm{P_{C}},\bm{P_{D}}}\quad &\sum\limits_{i\in\mathcal{C}}\sum\limits_{j\in\mathcal{D}} \left[\beta_{i,j}v_{i,j}(P_{C_{ij}},P_{D_{ji}})\right]-\gamma\delta(\bm
B)\\
\label{eq:probeq_Bconst}
\text{subject
to}\quad&\beta_{i,j}\in \{0,1\},~\forall
~i,j,~\sum\limits_{j=1}^{N_D}\beta_{i,j}\leq1 \,~\forall
~i,\\
&0\leq P_{C_{ij}}\leq P_{C_{max}},~~0\leq P_{D_{ji}}\leq P_{D_{max}},~\forall ~i,j,\\
&\dfrac{P_{C_{ij}}g_{C_i}}{N_0+P_{D_{ji}}h_{D_{j,i}}}\geq \eta^C_{min},~\dfrac{P_{D_{ji}}g_{D_j}}{N_0+P_{C_{ij}}h_{C_j}}\geq \eta^D_{min},~\forall ~i,j\label{eq:probeq_pconst}
\end{align}
\end{subequations}
where
\begin{myitemize}
\myitem $\bm{P_C}\triangleq [P_{C_{ij}}]$.
\end{myitemize}%
\end{myitemize}

\cmt{1: Power optimization}
\begin{myitemize}
\myitem Since $\bm B$ is binary,
 \eqref{eq:probeq} can now be decoupled without loss of optimality into a power allocation problem and a channel allocation problem. Furthermore, the optimization of \eqref{eq:probeq} with respect to $\bm{P_{C}}$ and $\bm{P_{D}}$ (power allocation problem)
decouples across $i$ and $j$ into the $N_C N_D$ sub-problems of the form:
\begin{align}
\label{eq:prob1}
\maximize_{P_{C_{ij}},P_{D_{ji}}}\quad & v_{i,j}(P_{C_{ij}},P_{D_{ji}})
\\\nonumber
\text{subject to } \quad & 0\leq P_{C_{ij}}\leq P_{C_{max}},\quad
0\leq P_{D_{ji}}\leq P_{D_{max}},
\\\nonumber 
&\dfrac{P_{C_{ij}}g_{C_i}}{N_0+P_{D_{ji}}h_{D_{j,i}}}\geq \eta^C_{min},\quad
\dfrac{P_{D_{ji}}g_{D_j}}{N_0+P_{C_{ij}}h_{C_j}}\geq \eta^D_{min},
\end{align}
which should be solved $\forall i\in\mathcal{C},\forall
j\in\mathcal{D}$. This power allocation subproblem coincides with the
one arising in~\cite{feng2013device,gjendemsjo2006optimal,elnourani2018underlay}, which can be solved in
closed-form, since the solution should be on the borders of the feasibility region (defined by the constraints in \eqref{eq:prob1}). 
More specifically, as illustrated in Fig. \ref{fig:PF}, it can be shown that for any point $(P'_{C_{ij}},P'_{D_{ji}})$ in the interior of the feasibility region, there exist a point $(\alpha P'_{C_{ij}}, \alpha P'_{D_{ji}})$ at the border segments that has a higher objective value. Moreover, since the objective function is convex on the border segments, and therefore the optimal point is one of the intersection points of the border segments (the maximum of a convex function is achieved at the borders of the feasibility region). 
\setlength{\textfloatsep}{0pt}
\begin{figure}[t]
\centering
\begin{tikzpicture}
\draw[name path = B] (1,0) -- (3,4);
\draw[name path = C] (0,1) -- (4,3);
\draw[name path = D] (2,2) -- (3,4) -- (4,4);
\draw[name path = E] (2,2) -- (4,3) -- (4,4);
\draw[line width=1.2mm] (3,4) -- (4,4);
\draw[line width=1.2mm] (4,3) -- (4,4);
\fill[gray, intersection segments={of=D and E, sequence={L2--R2}}];
\draw[name path = A] (0,0) rectangle (4,4);
\draw[dashed,-{Latex[length=3mm]}, to path={-| (\tikztotarget)}] (2.8,3.0) -- (2.8*4/3,4);
\draw[dashed,->, to path={-| (\tikztotarget)}] (0,0) -- (2.8*4/3,4);
\draw[->, to path={-| (\tikztotarget)}] (0,0) -- (0,5);
\draw[->, to path={-| (\tikztotarget)}] (0,0) -- (5,0);
\node (Py) at (0, 5.3) {$P_{D_{ji}}$};
\node (Px) at (5.4, 0) {$P_{C_{ij}}$};
\node (Py) at (-0.4, 4) {$P^D_{\max}$};
\node (Px) at (4, -0.3) {$P^C_{\max}$};
\node (P1) at (3, 4) {o};
\node (P1) at (3, 4.3) {$P_1$};
\node (P2) at (4, 4) {o};
\node (P2) at (4, 4.3) {$P_2$};
\node (P3) at (4, 3) {o};
\node (P3) at (4.3, 3) {$P_3$};
\node (P1) at (2.8, 3) {x};
\node (P1) at (2.8*4/3, 4) {x};
\draw[-{Latex[length=3mm]}, to path={-| (\tikztotarget)}] (1,3) -- (2.8,3);
\draw[-{Latex[length=3mm]}, to path={-| (\tikztotarget)}] (1,4.5) -- (2.8*4/3,4);
\node (P1) at (1, 3.2) {$(P'_{D_{ji}},P'_{C_{ij}})$};
\node (P2) at (1.25, 4.7) {$(\alpha P'_{D_{ji}},\alpha P'_{C_{ij}})$};
\end{tikzpicture}
\vspace{-5mm}
\caption{Power feasibility region}
\label{fig:PF}
\vspace{-5mm}
\end{figure}
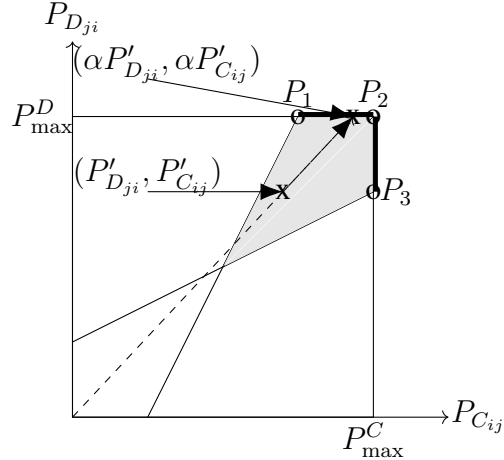
\myitem\cmt{2: Channel assignment}
\begin{myitemize}
\myitem\cmt{problem}
Once \eqref{eq:prob1} has been solved $\forall i\in\mathcal{C},\forall
j\in\mathcal{D}$, it remains to substitute the optimal values $[v^*_{i,j}]_{i,j}$ into \eqref{eq:probeq} and then minimize with respect to $\bm
B$. If \eqref{eq:prob1} is infeasible for a given $(i,j)$, then we set
its optimal value to $v^*_{i,j}=-\infty$.  The resulting channel
assignment subproblem can be expressed as follows,
\begin{align}
\label{eq:prob2}
\maximize\limits_{\bm
B}\quad& \sum_{i\in\mathcal{C}}\sum\limits_{j\in\mathcal{D}}
\beta_{i,j}v^*_{i,j}-\gamma\delta(\bm B),\\
\text{subject to}\quad&\beta_{i,j}\in \{0,1\}~\forall i,j, \quad
\sum\limits_{j\in\mathcal{D}}\beta_{i,j}\leq1 ~\forall i \nonumber.
\end{align}

\myitem\cmt{too difficult\ra suboptimality}Notice that problem \eqref{eq:prob2} is
an integer program of combinatorial nature. Finding an exact
solution using exhaustive search would be computationally unaffordable and time consuming for
a sufficiently large $N_C N_D$. Thus, considering the practicality of implementation, we compute a sub-optimal solution with a smaller computational complexity
by relaxing the integer constraint $\beta_{i,j}\in \{0,1\}$ of \eqref{eq:prob2} with
$\beta_{i,j}\in [0,1]$.

The resulting problem is convex and the resulting solutions $[\tilde \beta_{i,j}]$ can be efficiently obtained 
e.g. through projected gradient descent (PDG) \cite{bertsekas1999}.
\myitem\cmt{recover discrete solution} Discretizing the solution $[\tilde \beta_{i,j}]$
to such a problem is expected to yield an approximately optimal
optimum of \eqref{eq:prob2}.  For this discretization we consider two
approaches:
\begin{myitemize}%
\myitem\cmt{Maximum Mode}(i) for every $i$, set $\beta_{i,j}=1$ if
$j=\argmax_j \tilde \beta_{i,j}$.
\myitem\cmt{Random mode}(ii) for each $i$, consider a random variable
$J_i$ taking values $1,\ldots,N_D$ with probabilities $P(J_i=j)
= \tilde \beta_{i,j}/\sum\limits_{j\in\mathcal{D}}\tilde{\beta}_{i,j}$. 
Then, we generate a certain set of realizations of $\{J_i\}$ and
form the corresponding set of  matrices $\{\bm B$\}, whose $(i,j)$-th entry is 1 if $J_i=j$ and 0
otherwise. Finally, we evaluate the objective of \eqref{eq:prob2} for all
these realizations and select the realization with the highest
objective value.
\end{myitemize}

\acom{Another way to relax the problem is modifying the binary constraints to $\beta_{i,j}(1-\beta_{i,j})\leq 0$. The Lagrangian of this problem can be expressed as: $L(\bm B, \bm \Lambda, z)=\sum_{i\in\mathcal{C}}\left(\sum\limits_{j\in\mathcal{D}}
[\beta_{i,j}v_{i,j}-\lambda_{ij}\beta_{i,j}(1-\beta_{i,j})]-z_i(\bm B_i\bf 1)\right)-\gamma\delta^2(\bm B)$}
\end{myitemize}
\end{myitemize}
\vspace{-4mm}
\subsection{Resource Allocation Under Imperfect \gls{CSI} (ICSI)}
\label{subsec:preRand}
\cmt{Criterion}
\begin{myitemize}
\myitem\cmt{Scenario}
In this scenario, we assume having infrequent and limited measurements from the \gls{CUs} and the D2D pairs that are used in estimating the channel gain from the D2D pairs to the CUs. This will create uncertainty in the available CSI. 
Thus, in this case, the objective function and the SINR constraints in \eqref{eq:probeq} involve a random channel gain $\tilde{h}_{D_{j,i}}$ for the interference link from the $j$-th \gls{DtD} pair to the $i$-th \gls{CU}.
\par
First, the SINR constraint \eqref{eq:probeq_pconst} can be replaced with a  probabilistic constraint to guarantee a maximum outage probability $\epsilon$ which can be expressed as:
\myitem\cmt{overall optimization problem}
\begin{align}
   \Pr\left\{\eta_{ij}^c\triangleq\dfrac{P_{C_{ij}}g_{C_i}}{N_0+P_{D_{ji}}\tilde{h}_{D_{j,i}}}\geq \eta^c_{min}\right\}\geq 1-\epsilon.
   \label{eq:pCon}
\end{align}
\end{myitemize}
\cmt{proposed solution}
\begin{myitemize}
\myitem The probabilistic constraint in \eqref{eq:pCon} can be expressed in closed form for a given statistical distribution of $\tilde{h}_{D_{ji}}$. 
\myitem Generally, \eqref{eq:pCon} is equivalent to:
$\Pr\left\{\tilde{h}_{D_{j,i}}\leq \dfrac{P_{C_{ij}}g_{C_i}-\eta^c_{min}N_0}{P_{D_{ji}}\eta^c_{min}}\right\}\geq 1-\epsilon $, 
or, equivalently
\begin{align}
\label{eq:cons_CF}
&\dfrac{P_{C_{ij}}g_{C_i}}{N_0+P_{D_{ji}}\text{F}^{-1}_{\tilde{h}_{D_{j,i}}}(1-\epsilon)}\geq \eta^c_{min},
\end{align}
where $\text{F}^{-1}_{\tilde{h}}(1-\epsilon)$ is the inverse cumulative distribution function (CDF) function for $\tilde{h}$ evaluated at $1-\epsilon$. We will consider exponential, Gaussian, Chi-squared, and log-normal distributions in the following sections, since they are the most common in wireless communication environment.
\myitem{\cmt{The objective function}\acom{\\}}

Next, focussing on the objective function, we consider two approaches: (i) expected network rate maximization; and (ii) minimum network rate maximization. 

\subsubsection{Expected Network Rate Maximization (ERM)}
One possibility is to replace the objective of \eqref{eq:prob} with its expectation. To this end, notice that $\mathbb{E}_{\bold{\tilde{h}}}\{R\} = \sum_{i\in\mathcal{C}}\left[\sum_{j\in\mathcal{D}} \beta_{i,j}\mathbb{E}_{\tilde{h}_{D_{j,i}}} \{v_{i,j}\} + R_{C_{i,0}} \right]$, and from the definition of $v_{ij}$:
\begin{align*}
&\mathbb{E}_{\tilde{h}_{D_{j,i}}}\{v_{i,j}(P_{C_{ij}},P_{D_{ji}})\}=\mathbb{E}_{\tilde{h}_{D_{j,i}}}\{R_{C_{i,j}}(P_{C_{ij}},P_{D_{ji}})\} 
+R_{D_{j,i}}(P_{C_{ij}},P_{D_{ji}})-R_{C_{i,0}}.
\end{align*}
Since the expectation of $R_{C_{i,j}}(P_{C_{ij}},P_{D_{ji}})$ is not tractable analytically for the aforementioned distributions, one can replace the expectation by the first-order or the second-order Taylor series approximations around the mean of ${\tilde{h}_{D_{j,i}}}$.
\setlength{\textfloatsep}{0pt}
\begin{table}[t]
    \centering
    \begin{tabular}{|c||p{3.5cm}|p{3.5cm}|} \hline
         Calculation method & \multicolumn{2}{p{7cm}|}{Deviation $|\hat{v}_{i,j}-\bar{v}_{i,j}|/\bar{v}_{i,j}$ between the approximation $\hat{v}_{i,j}$ and the Monte-Carlo average $\bar{v}_{i,j}$ for $10^6$ samples} \\ \hline
         \backslashbox{Distributions}{Approximation} & First order & Second order \\\hline\hline
         Exponential ($\bar{h}_{C_{ij}}=0.2$) & 0.6499\% & 0.1392\% \\\hline
         Gaussian ($\bar{h}_{C_{ij}}=0.2,~\var{h_{C_{ij}}}=0.01$) & 0.8934\% & 0.1062\% \\\hline
         Chi-squared ($\bar{h}_{C_{ij}}=0.2,~\var{h_{C_{ij}}}=0.01$)& 0.8930\% & 0.1058\%\\\hline
         Log-normal ($\bar{h}_{C_{ij}}=0.2,~\var{h_{C_{ij}}}=0.01$)& 0.6898\% &  0.0991\% \\\hline
    \end{tabular}
    \vspace{-5mm}
    \caption{Expectation approximations}
    \label{tab:TayCom}
\end{table}
Table \ref{tab:TayCom} shows the comparison between first-order and second-order approximations in the computation of $\mathbb{E}_{\tilde{h}_{D_{j,i}}}\{v_{ij}\}$, where we can note that both approximations are very close to the Monte-Carlo averages in all the tested distributions. 
Besides, the first-order approximation results in an error comparable to the second-order approximation. Because of this reason and the higher simplicity, we consider the first-order approximation.
Moreover, the resulting expectation is the so-called certainty equivalence approximation, which is an extensively adopted approximation in stochastic optimization \cite{bertsekas1999}.
Using the expectation of the first-order Taylor approximation in the objective function along with aforementioned constraints in \eqref{eq:cons_CF} leads to a problem similar to \eqref{eq:prob1}, which can be solved in closed form as before.
\subsubsection{Minimum guaranteed rate maximization (MRM)}
In this approach, the criterion to maximize is the network rate 
exceeded 
for  a (1-$\epsilon$) portion of the time. First, we define the $(1-\epsilon)$-guaranteed SINR for the $i$-th CU when sharing the channel with the $j$-th D2D pair as $\eta^\epsilon_{C_{i,j}}$ such that $\text{Pr}\{\eta_{C_{i,j}}>\eta_{C_{i,j}}^{\epsilon}\}=1-\epsilon$. Next, we define $v^\epsilon_{ij}=\log(1+\eta^\epsilon_{C_{i,j}})+\log(1+\eta_{D_{i,j}})-R_{C_{i,0}}$ and $R^\epsilon = \sum_{i\in\mathcal{C}}\left[\sum_{j\in\mathcal{D}} \beta_{i,j}v^\epsilon_{i,j} + R_{C_{i,0}} \right]$, siilar to the work in \cite{elnourani2019reliable}. The resource allocation problem can be formulated as \eqref{eq:prob} with $R$ replaced by $R^\epsilon$ and the SINR constraints replaced by \eqref{eq:cons_CF}. Proceeding as in Sec. \ref{subsec:preD}, such a problem is equivalent to \eqref{eq:probeq} with $v_{ij}$ replaced with $v^\epsilon_{ij}$ and the SINR constraints replaced by \eqref{eq:cons_CF}. Similar steps can also be followed to decouple the problem into power assignment and channel allocation subproblems. Up to a constant term, the objective of the power allocation sub-problems can be expressed as:
\begin{align}
\label{eq:d2r1}
F_0\triangleq \log_2\left(1+\dfrac{P_{C_{ij}}g_{C_i}}{N_0+P_{D_{ji}}\text{F}^{-1}_{\tilde{h}_{{D_{j,i}}}}(1-\epsilon)}\right)
+\log_2\left(1+\dfrac{P_{D_{ji}}g_{D_j}}{N_0+P_{C_{ij}}h_{C_{j}}}\right)
\end{align}
where $\eta^\epsilon_{C_{i,j}}$ is expressed in closed-form similar to \eqref{eq:pCon} for a given statistical distribution of $\tilde{h}_{D_{j,i}}$. The rest of this section proposes a method to solve this power allocation subproblem.

\myitem{\cmt{Fractional Programming. }}
This objective function is non-convex. However, it can be seen as a sum of log-functions of ``concave-over-convex" fractions. 
Given this structure, fractional programming techniques \cite{shen2018fractional,shen2018fractional2} constitute a natural fit.
To take the fractions outside the log-functions, we introduce the slack variables $\bold{z} \triangleq [z_1,~z_2]^T$. 
The resulting power assignment problem can be rewritten as follows:
\begin{subequations}
\label{eq:d2r}
\begin{align}
\maximize_{P_{C_{ij}},P_{D_{ji}},\bm{z}}\quad 
&\log_2\left(1+z_1\right)+\log_2\left(1+z_2\right) 
\\\label{eq:d2rz}
\text{subject to } \quad & z_1\leq\dfrac{P_{C_{ij}}g_{C_i}}{N_0+P_{D_{ji}}\text{F}^{-1}_{\tilde{h}_{D_{j,i}}}(1-\epsilon)},\quad
z_2\leq\dfrac{P_{D_{ji}}g_{D_j}}{N_0+P_{C_{ij}}h_{C_{j}}}
\\\label{eq:d2rmaxp}
&0\leq P_{C_{ij}}\leq P_{C_{max}},\quad
0\leq P_{D_{ji}}\leq P_{D_{max}},
\\\label{eq:d2rSINR}
&\dfrac{P_{C_{ij}}g_{C_i}}{N_0+P_{D_{ji}}\text{F}^{-1}_{\tilde{h}_{D_{j,i}}}(1-\epsilon)}\geq \eta^C_{min},\quad
\dfrac{P_{D_{ji}}g_{D_j}}{N_0+P_{C_{ij}}h_{C_j}}\geq \eta^D_{min}.
\end{align}
\end{subequations}
The optimal values of the auxiliary variables occur when the inequalities hold with equality ($z_1^*={P_{C_{ij}}g_{C_i}}/{N_0+P_{D_{ji}}\text{F}^{-1}_{\tilde{h}_{D_{j,i}}}(1-\epsilon)},~ z_2^*={P_{D_{ji}}g_{D_j}}/{N_0+P_{C_{ij}}h_{C_{j}}}$). 
Let us consider the Lagrangian of \eqref{eq:d2r} with respect to the first two inequalities:
\begin{align}
\label{eq:d2rs}
L(\bm p,\bm{z},\bm\lambda)&= \log_2\left(1+z_1\right)+\log_2\left(1+z_2\right)
- \lambda_1\left(z_1-\dfrac{P_{C_{ij}}g_{C_i}}{N_0+P_{D_{ji}}\text{F}^{-1}_{\tilde{h}_{D_{j,i}}}(1-\epsilon)}\right)\nonumber\\ &
-\lambda_2\left(z_2-\dfrac{P_{D_{ji}}g_{D_j}}{N_0+P_{C_{ij}}h_{C_{j}}}\right)
\end{align}
A stationary point of $L$ with respect to $\bm{z}$ is achieved when ${\partial L}/{\partial \bm z}=\bm 0$. This leads to $\lambda_1=1/(1+z_1),~\lambda_2=1/(1+z_2)$. Substituting $\bm z^*$ 
in these equations yields:
\begin{align}
\label{eq:d2v}
\lambda_1^*= 
\dfrac{N_0+P_{D_{ji}}\text{F}^{-1}_{\tilde{h}_{D_{j,i}}}(1-\epsilon)}{P_{C_{ij}}g_{C_i}+N_0+P_{D_{ji}}\text{F}^{-1}_{\tilde{h}_{D_{j,i}}}(1-\epsilon)},
\quad \lambda_2^*=
\dfrac{N_0+P_{C_{ij}}h_{C_{j}}}{P_{D_{ji}}g_{D_j}+N_0+P_{C_{ij}}h_{C_{j}}}
\end{align}
Substituting $\mathbf{\lambda}^*$ in \eqref{eq:d2rs}, we obtain:
\begin{align}
\label{eq:d4}
\maximize_{P_{C_{ij}},P_{D_{ji}},\bm{z}}\quad &F_1\triangleq L(\bm p,\bm{z},\bm\lambda^*)=  \log_2\left(1+z_1\right)
+\log_2\left(1+z_2\right)-z_1  \\\nonumber
&+\dfrac{(1+z_1)P_{C_{ij}}g_{C_i}}{P_{C_{ij}}g_{C_i}+N_0+P_{D_{ji}}\text{F}^{-1}_{\tilde{h}_{D_{j,i}}}(1-\epsilon)}-z_2
+\dfrac{(1+z_2)P_{D_{ji}}g_{D_j}}{P_{D_{ji}}g_{D_j}+N_0+P_{C_{ij}}h_{C_{j}}}
\\\nonumber
\text{subject to } \nonumber\quad&\eqref{eq:d2rmaxp}, ~\eqref{eq:d2rSINR} 
\end{align}
Finally, to handle the fractions in the objective function, we use the quadratic transformation in \cite{shen2018fractional,shen2018fractional2}, to transform each fraction into a substitute concave expression. 
Then, we obtain:
\begin{align}
\label{eq:d5}
\maximize_{P_{C_{ij}},P_{D_{ji}},\bm{z},\bm y}\quad &F_2\triangleq   \log_2\left(1+z_1\right)+\log_2\left(1+z_2\right)  
-z_1+2y_1\sqrt{(1+z_1)P_{C_{ij}}g_{C_i}}\\\nonumber
&-y_1^2(P_{C_{ij}}g_{C_i}+N_0+P_{D_{ji}}\text{F}^{-1}_{\tilde{h}_{D_{j,i}}}(1-\epsilon))
-z_2+2y_2\sqrt{(1+z_2)P_{D_{ji}}g_{D_j}}\\\nonumber &
-y_2^2(P_{D_{ji}}g_{D_j}+N_0+P_{C_{ij}}h_{C_{j}})
\\\nonumber
\text{subject to } \quad &\eqref{eq:d2rmaxp}, ~\eqref{eq:d2rSINR},
\end{align}
where $\mathbf{y} \triangleq [y_1 ~ y_2]^T$ are the auxiliary variables given by the quadratic transformation.

This problem is then solved by alternating maximization with respect to the individual $y_1,y_2$, $P_{C_{ij}},P_{D_{ji}}$ variables. At each step, all iterates can be 
obtained in closed form by taking the partial derivative with respect to each variable and setting it to $0$, and projecting the solution onto the feasible set. The overall iteration can be expressed as:
{\small\begin{subequations}
\label{eq:FP_sol}
\begin{align}
\label{eq:sol_zs}
z_1^{[k+1]}=&\dfrac{P_{C_{ij}}^{[k]}g_{C_i}}{N_0+P_{D_{ji}}^{[k]}\text{F}^{-1}_{\tilde{h}_{D_{j,i}}}(1-\epsilon)},~ \quad z_2^{[k+1]}=\dfrac{P_{D_{ji}}^{[k]}g_{D_j}}{N_0+P_{C_{ij}}^{[k]}h_{C_{j}}}\\
\label{eq:sol_ys}
y_1^{[k+1]} &=\dfrac{\sqrt{(1+z_1^{[k+1]})P_{C_{ij}}^{[k]}g_{C_i}}}{P_{C_{ij}}^{[k]}g_{C_i}+N_0+P_{D_{ji}}^{[k]}\text{F}^{-1}_{\tilde{h}_{D_{j,i}}}(1-\epsilon)},~
 y_2^{[k+1]} =\dfrac{\sqrt{(1+z_2^{[k+1]})P_{D_{ji}}^{[k]}g_{D_j}}}{P_{D_{ji}}^{[k]}g_{D_j}+N_0+P_{C_{ij}}^{[k]}h_{C_{j}}}
\\\label{eq:pow_FP1} P_{C_{ij}}^{[k+1]}&= \text{Proj}_{\mathcal{S}_1^{[k]}}\left(\dfrac{(y_1^{[k+1]})^2(1+z_1^{[k+1]})g_{C_i}}{((y_1^{[k+1]})^2g_{C_i}+(y_2^{[k+1]})^2h_{C_{j}})^2}\right)
\\\label{eq:pow_FP2}
P_{D_{ji}}^{[k+1]}&=\text{Proj}_{\mathcal{S}_2^{[k+1]}}\left( \dfrac{(y_2^{[k+1]})^2(1+z_2^{[k+1]})g_{D_j}}{((y_2^{[k+1]})^2g_{D_j}+(y_1^{[k+1]})^2 \text{F}^{-1}_{\tilde{h}_{D_{j,i}}}(1-\epsilon))^2}\right)
\end{align}
\end{subequations}}
where $k$ is the iteration index, $\text{Proj}_{\mathcal{A}}(\bm *)$ is a projection of $\bm *$ onto the set $\mathcal{A}$; $S_1^{[k]}\triangleq \{P_{C_{ij}}:(P_{C_{ij}},P_{D_{ji}}^{[k]})$ satisfy \eqref{eq:d2rmaxp} and \eqref{eq:d2rSINR}\}, 
and $S_2^{[k+1]}\triangleq \{P_{D_{ji}}:(P_{C_{ij}}^{[k+1]},P_{D_{ji}})$ satisfy \eqref{eq:d2rmaxp} and \eqref{eq:d2rSINR}\}. 
Next, we show that, with this alternating optimization solution, $\vert F_0(P_{C_{ij}}^{[k]},P_{D_{ji}}^{[k]})-F_0(P_{C_{ij}}^{*},P_{D_{ji}}^{*})\vert$ converges in the order $\mathcal{O}(k^{-\alpha})$, for some $\alpha>0$).

\end{myitemize}
\begin{theorem}
Let $\{\mathbf{p}^{[k]}\}_{k \in \mathbb{N}_+}$ be the sequence generated by \eqref{eq:FP_sol} with $\mathbf{p}^{[k]} \triangleq [P_{D_{ji}}^{[k]} ~ P_{C_{ij}}^{[k]}]^T$. Then, (i) $\lim\limits_{k\rightarrow\infty} \mathbf{p}^{[k]}=\bar{\mathbf{p}}$ where $\bar{\mathbf{p}}$ is a stationary point of \eqref{eq:d5}, and (ii) $\vert \mathbf{p}^{[k]}-\bar{\mathbf{p}}\vert\leq C k^{-(1-\theta)/(2\theta-1)}$ for some $C>0$.
\end{theorem}
\vspace{-5mm}
\begin{proof}
see Appendix \ref{App:Proof1}
\end{proof}
\acom{
Numerically, the KL property bound $\eta$ for 10,000 points located uniformly at random in a small region around the obtained stationary point $\bar{\mathbf p}$ at different values of $\theta$ is shown in fig. \ref{fig:AppBound}. It shows for that for $\theta>0.6$, $\eta$ can be considered bounded with great certainty. Fig. \ref{fig:AppConv} shows the convergence of the proposed algorithm which shows a sublinear convergence as proven.
\begin{figure}
\centering
\includegraphics[width=0.48\textwidth]{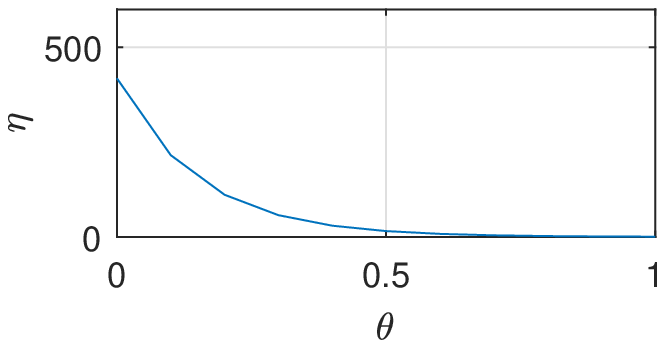}
\caption{\%The bound $\eta$ vs. $\theta$}
\label{fig:AppBound}
\vspace{-5mm}
\end{figure}
\begin{figure}
\centering
\includegraphics[width=0.48\textwidth]{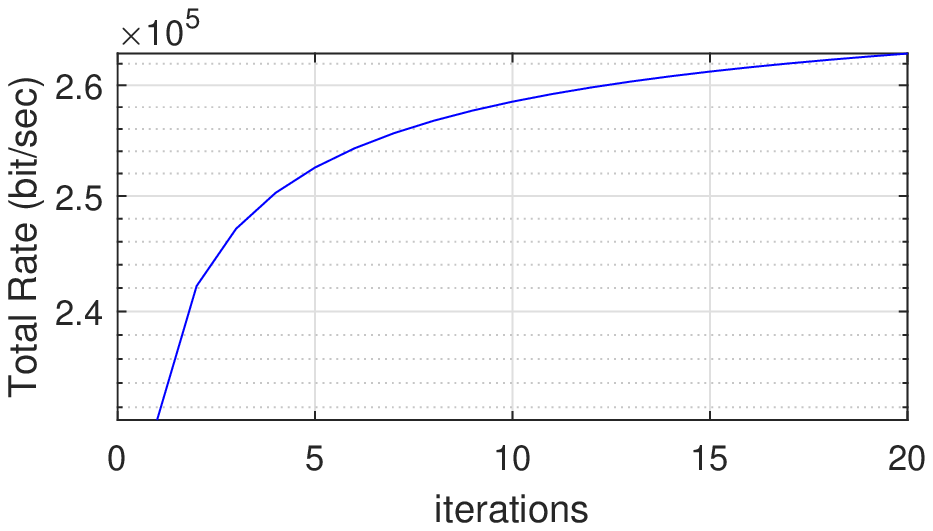}
\caption{\%Convergence of the proposed method}
\label{fig:AppConv}
\vspace{-5mm}
\end{figure}}

After solving the power allocation subproblems in both the ERM or MRM cases, a channel allocation problem similar to \eqref{eq:prob2} will arise, and a similar solution based on integer relaxation can be used. Algorithm \ref{algo:Cent_Ind} highlights the operation of the separate resource allocation method with all the previously discussed cases, with $\bm B^{(d)}(k),\bm B^{(u)}(k),\bm P_C^{(d)}(k),\bm P_D^{(d)}(k),\bm P_D^{(u)}(k),$ $\bm P_C^{(u)}(k)$ are the values of each variable at the $k$-th iteration, and $\bf P_{D_j}^{(d)},\bf P_{C_j}^{(d)}$ are the $j$-th columns of $\bf P_{D}^{(d)},\bf P_{C}^{(d)}$ matrices. Since the objective function of the channel allocation problem is Lipschitz smooth, this algorithm will converge as $\mathcal{O}(1/k)$ (as shown in Theorem 3.7 in \cite{bubeck2015convex}) in the case of PCSI and ICSI-ERM with $\mathcal{O}(N_C^2N_D)$ computational operations per iteration. Similarly, in the case of ICSI-MRM, the algorithm will converge as $\mathcal{O}(1/k+k^{-\alpha})$ for some $\alpha>0$, with similar computations per iteration.
\setlength{\textfloatsep}{0pt}
\begin{algorithm}[t] 
\caption{Centralized Separate Resource Allocation} 
\label{algo:Cent_Ind} 
\begin{algorithmic}[1] 
    \State 	Initialize: $\bm B^{(d)}(0),\bm B^{(u)}(0),\bm P_C^{(d)}(0),\bm P_D^{(d)}(0),\bm P_D^{(u)}(0),$ $\bm P_C^{(u)}(0),k=0$
    \ForAll{$j\in \mathcal{D}^{(d)}$}
    \Comment{Power Assignment in downlink to find: $\bf P_{D_j}^{(d)^*},\bf P_{C_j}^{(d)^*}$.}
    \If{PCSI mode \textbf{OR} ICSI-ERM mode} 
    \State BS uses the closed-form power allocation in Sec. \ref{subsec:preD}.
    \Else 
    \algorithmiccomment{\emph{ICSI-MRM mode}}
    \State BS applies \eqref{eq:FP_sol} iteratively until convergence.
    \EndIf
    \EndFor
    \Repeat
    \State $k=k+1$
    \State 	BS uses the PGD algorithm to calculate: $\bm B^{(d)}(k)$
    \Until{$\bm B^{(d)}$ converges}
    \State \dots \Comment{The same for uplink}
    \State BS discretize $\bm B^{(d)},\bm B^{(u)}$.
\end{algorithmic}
\end{algorithm}
\vspace{-5mm}
\section{Joint Uplink and Downlink Resource Allocation}
\label{sec:JointUD}
In this section, we analyze the scenario in which \gls{DtD}  pairs are assigned uplink or downlink channels on the basis of instantaneous channel conditions, i.e., the algorithm itself generates a decision on the set of \gls{DtD} pairs communicating in the uplink or the downlink while maximizing the aggregate network throughput.\begin{myitemize}
\myitem\cmt{decouples into two similar sub-problems}Problem \eqref{eq:probeq} can be extended to the joint uplink and downlink resource allocation case by considering the following modified objective function:
\begin{align}\label{eq:probext}
&\maximize\limits_{\bm B,\bm{P_{C}},\bm{P_{D}}} \sum\limits_{i\in\mathcal{C}}\sum\limits_{j\in\mathcal{D}} [\beta^{(u)}_{i,j}U^{(u)}_{i,j}(P^{(u)}_{C_{ij}},P^{(u)}_{D_{ji}})
+\beta^{(d)}_{i,j}U^{(d)}_{i,j}(P^{(d)}_{C_{ij}},P^{(d)}_{D_{ji}})]-\gamma\delta_J(\bm B^{(u)},\bm B^{(d)}),
\end{align}
where $U^{(u)}_{i,j}(P^{(u)}_{C_{ij}},P^{(u)}_{D_{ji}})$ and $U^{(d)}_{i,j}(P^{(d)}_{C_{ij}},P^{(d)}_{D_{ji}})$ are general utility functions for the uplink and downlink selected depending on the working conditions (PCSI, ICSI-ERM, or ICSI-MRM), which are set to either the rate gain or the expected rate gain or the minimum rate gain defined in Sec. \ref{sec:SepDU}. In addition, the constraints must also be extended to take into account both up-link and down-link communications.

\par
\myitem Here, we redefine a joint unfairness metric $\delta_J$ for joint resource allocation in uplink and downlink. 
\myitem Let $N_D$ be the number of D2D pairs and let
\myitem $N_C^{(u)}$ and $N_C^{(d)}$ be the total number of channels available in the uplink and downlink respectively.
\myitem A D2D pair is allowed to communicate in either the downlink or uplink ($\sum\limits_{i \in \mathcal{C}^{(d)}}\beta^{{(d)}}_{i,j}\times\sum\limits_{i \in \mathcal{C}^{(u)}}\beta^{{(u)}}_{i,j}=0~\forall j$).
\myitem The fairest possible assignment is the one assigning $m_0=(N_C^{(u)}+N_C^{(d)})/N_D=N_C^{(u)}/N_D+N_C^{(d)}/N_D\triangleq m_0^{(u)}+m_0^{(d)}$ to each D2D pair. Similarly to Sec. \ref{sec:Sys_Mod}, we can adopt the following fairness metric:
{\small
\begin{align*}
&\delta_J(\bm B^{{(u)}},\bm B^{{(d)}})=\dfrac{\sum\limits_{j \in \mathcal{D}} (m_j-m_0)^2}{m_0^2~N_D} 
=\dfrac{1}{m_0^2~N_D} \sum\limits_{j \in \mathcal{D}} \left(\sum\limits_{i \in \mathcal{C}^{(u)}}\beta^{{(u)}}_{i,j}+\sum\limits_{i \in \mathcal{C}^{(d)}}\beta^{{(d)}}_{i,j}-m_0^{(u)}-m_0^{(d)}\right)^2
\end{align*}
\begin{align}\nonumber
&=\dfrac{(m_0^{(u)})^2}{m_0^2}\delta(\bm B^{{(u)}})-\dfrac{2}{m_0^2~N_D} \sum\limits_{j \in \mathcal{D}} \left(m_0^{(d)} \sum\limits_{i \in \mathcal{C}^{(u)}}\beta^{{(u)}}_{i,j}\right)
+\dfrac{(m_0^{(d)})^2}{m_0^2}\delta(\bm B^{{(d)}})\\&
-\dfrac{2}{m_0^2~N_D} \sum\limits_{j \in \mathcal{D}} \left(m_0^{(u)} \sum\limits_{i \in \mathcal{C}^{(d)}}\beta^{{(d)}}_{i,j}\right)
+\dfrac{2 m_0^{(u)}m_0^{(d)}}{m_0^{2}}
\end{align}}
\myitem The resulting optimization problem is now given by:

{\small
\begin{subequations}
\label{eq:probj}
\begin{align} 
\maximize\limits_{\bm B^{{(u)}},\bm B^{{(d)}},\bm P^{{(u)}},\bm P^{{(d)}}}& \sum\limits_{i\in\mathcal{C}}\sum\limits_{j\in\mathcal{D}} \left[\beta^{{(u)}}_{i,j}U^{{(u)}}_{i,j}(P^{{(u)}}_{C_{ij}},P^{{(u)}}_{D_{ji}})
+\beta^{{(d)}}_{i,j}U^{{(d)}}_{i,j}(P^{{(d)}}_{C_{ij}},P^{{(d)}}_{D_{ji}})\right]
-\gamma\delta_J(\bm B^{{(u)}},\bm B^{{(d)}}),\\\label{eq:jc1}
\text{s.t.}~ &\beta^{{(u)}}_{i,j}\in \{0,1\},~~\beta^{{(d)}}_{i,j}\in \{0,1\},\quad~\forall ~i,j\\\label{eq:jc2}
&\sum\limits_{j\in \mathcal{D}}\beta^{{(d)}}_{i,j}\leq1, ~~
\sum\limits_{j\in\mathcal{D}}\beta^{{(u)}}_{i,j}\leq1, \quad~\forall ~i\\\label{eq:jc3}
&\left(\sum\limits_{i \in \mathcal{C}^{(d)}}\beta^{{(d)}}_{i,j}\right)\times\left(\sum\limits_{i \in \mathcal{C}^{(u)}}\beta^{{(u)}}_{i,j}\right)=0,\quad~\forall ~j\\
&\text{UL/DL power and SINR constraints similar to } \eqref{eq:probeq}
\end{align}
\vspace{-2mm}
\end{subequations}}
\myitem Similar to \eqref{eq:probeq}, this problem can further be decomposed into power and channel problems as before without loss of optimality. 
\myitem The power allocation problems are of the form of \eqref{eq:prob1} or \eqref{eq:d2r} depending on the available CSI (PCSI or ICSI) and the selected criteria (ERM or MRM).
\end{myitemize}
\vspace{-4mm}
\subsection{Resource Allocation under Perfect \gls{CSI} (PCSI)}
\label{subsec:clasDet}
In this case, the objective function in \eqref{eq:probj} becomes deterministic and the decomposition of the problem leads to a similar independent power allocation problem for each pair ($i,j$) as in \ref{subsec:preD}, which can be solved in closed form and obtain the optimal $U^{{(u)}^*}_{i,j}, ~U^{{(d)}^*}_{i,j}$.\\
The channel allocation problem becomes:
\begin{subequations}
\label{eq:probextch}
\begin{align} 
\maximize\limits_{\bm B^{{(u)}},\bm B^{{(d)}}}& \sum\limits_{i\in\mathcal{C}}\sum\limits_{j\in\mathcal{D}} [\beta^{{(u)}}_{i,j}U^{{(u)}^*}_{i,j}
+\beta^{{(d)}}_{i,j}U^{{(d)}^*}_{i,j}]
-\gamma\delta_2(\bm B^{{(u)}},\bm B^{{(d)}}),\\\label{eq:binConsJ}
\text{s.t.}~ &\eqref{eq:jc1},~\eqref{eq:jc2},~\text{and}~\eqref{eq:jc3}.
\end{align}
\end{subequations}
Relaxing the problem by ignoring the constraints \eqref{eq:jc3} and converting the binary constraints in \eqref{eq:jc1} to linear constraints as in sec. \ref{subsec:preD}, leads to a convex problem with linear constraints. This problem can also be solved using PGD, since it is differentiable with linear constraints. 
Finally, the obtained solution needs to be discretized and projected onto to the set defined by \eqref{eq:jc3}. We propose obtaining a binary solution in the same way used for discretizing \eqref{eq:prob2}. Afterwards, for each pair, we evaluate the objective function with the pair assigned to either uplink or downlink; we then select the one which has a higher objective function. After that, we then repeat the channel assignment with the pair removed from the deselected spectrum. The whole process is then repeated until all pairs are assigned. In general, there are many ways to project a solution into  the constraints in \eqref{eq:jc3}, however, one cannot guarantee optimallity since \eqref{eq:probextch} has been relaxed. 
\vspace{-5mm}
\subsection{Resource Allocation under Imperfect \gls{CSI} (ICSI)}
\label{subsec:clasRand}
The power allocation subpoblems here will be similar to Sec. \ref{subsec:preRand} and will adhere to similar solutions.
The channel allocation problem is similar to Sec. \ref{subsec:clasDet} and will follow the same solutions.
Algorithm \ref{algo:Cent_J} describes the operation of the joint resource allocation methods. As shown in Sec. \ref{sec:SepDU}, this algorithm will converge as $\mathcal{O}(1/k)$ in the case of PCSI and ICSI-ERM, and in the case of ICSI-MRM, the algorithm will converge as $\mathcal{O}(1/k+k^{-\alpha})$ for some $\alpha>0$, 
with $\mathcal{O}(N_C^2N_D)$ computational operations per iteration in all cases.
\setlength{\textfloatsep}{0pt}
\begin{algorithm}[t] 
\caption{Centralized Joint Resource Allocation} 
\label{algo:Cent_J} 
\begin{algorithmic}[1] 

    \State 	Initialize: $\bm B^{(d)}(0),\bm B^{(u)}(0),\bm P_C^{(d)}(0),\bm P_D^{(d)}(0),\bm P_D^{(u)}(0),$ $\bm P_C^{(u)}(0),k=0$\\ 
    \ForAll{$j\in \mathcal{D}$}
    \Comment{Power assignment in both downlink and uplink to find: $\bf P_{D_j}^{(d)^*},\bf P_{C_j}^{(d)^*},\bf P_{D_j}^{(u)^*},\bf P_{C_j}^{(u)^*}$.}
    \If{PCSI mode \textbf{OR} ICSI-ERM mode}
    \State BS uses the closed-form power allocation in Sec. \ref{subsec:preD}.
    \Else \Comment{ICSI-MRM mode}
    \State BS applies \eqref{eq:FP_sol} iteratively until convergence.
    \EndIf
    \EndFor
    \Repeat
    \State $k=k+1$
    \State 	BS uses the PGD algorithm to calculate: $\bm B^{(d)}(k),\bm B^{(u)}(k)$
    \Until{$\bm B^{(d)},\bm B^{(u)}$ converges}
    
    \State BS discretize $\bm B^{(d)},\bm B^{(u)}$.
\end{algorithmic}
\end{algorithm}
\vspace{-4mm}
\section{Decentralized Algorithms}
\label{sec:Dec}
In order to limit dependence of \gls{DtD} communication on \gls{BS} together with reducing BS's computational load, we also consider decentralizing the resource allocation algorithms. Furthermore, 
our aim is to start the communication immediately after  the  first iteration without waiting for convergence of the algorithm. Since the power assignment subproblems are independent, they can be solved entirely by the D2D pairs. To decompose the channel allocation problem, let $G$ be the objective function of \eqref{eq:prob2}. $G$ and its gradient can be express as:

\begin{subequations}
\label{eq:G}
\begin{align}
    &G=\text{vec}(\bm V) ^T\text{vec}(\bm B)-k_1 \Vert \bm B^T \bm 1 - k_2 \bm 1\Vert^2,\\
    &\nabla G=\bm V-2 k_1 (\bm 1 \bm 1^T \bm B- k_2\bm 1 \bm 1^T),
\end{align}
\end{subequations}
where $\bm V\triangleq [v_{ij}]$, and $k_1$ and $k_2$ are constants.
 Notice that, the channel allocation problem can not be directly decomposed into disjoint subproblems for each D2D pair, due to the quadratic term in \eqref{eq:G}. Nevertheless, the gradient is linear in $\bm B$, thus, the descent part of the channel allocation algorithm can be directly decomposed and each D2D pair can perform an optimization step over its corresponding part, without loss of optimality. Only the projection and the discretization have to be performed centrally at the \gls{BS}.
 \vspace{-5mm}
\subsection{Separate Uplink and Downlink Resource Allocation}
Algorithm \ref{algo:Dec_I} below describes how this scenario can be solved in a decentralized manner.
The BS initializes the power assignment vectors and the channel allocation matrices, and broadcasts them. Each D2D pair perform a step of the power allocation algorithm suitable for the network operation scenario (i.e. closed form for PCSI or ICSI-ERM or the alternating minimization in \eqref{eq:FP_sol} for ICSI-MRM). Then, each D2D pair updates its vectors of the channel allocations ($\bm B^{{(d)}}_j, \bm B^{{(u)}}_j$) by performing a gradient step. Each D2D pair then sends its channel allocation vectors along with the calculated power values to the BS. Then, the BS assembles all the vectors of the channel allocation matrices and projects them into a feasible solution and resends them to all D2D pairs. The BS and D2D pairs uses these calculated powers and channel assignments for communications. These steps are then repeated until all variables converge.
Algorithm \ref{algo:Dec_I} will also 
converge as $\mathcal{O}(1/k)$ in the case of PCSI and ICSI-ERM, and as $\mathcal{O}(1/k+k^{-\alpha})$ for some $\alpha>0$ for the case of ICSI-MRM. 
However, $\mathcal{O}(N_C^2)$ computational operations per iteration are performed by each D2D pair, and $\mathcal{O}(N_C N_D)$ computational operations per iteration are performed by the BS with $2 N_C N_D$ variables exchanged between the D2D pairs and the BS in every iteration.
\setlength{\textfloatsep}{0pt}
\begin{algorithm}[t] 
\caption{Decentralized Separate Resource Allocation} 
\label{algo:Dec_I} 
\begin{algorithmic}[1] 
    \State 	Initialize: $\bm B^{{(d)}}(0),\bm B^{{(u)}}(0),\bm P_C^{{(d)}}(0),\bm P_D^{{(d)}}(0),\bm P_D^{{(u)}}(0),$ $\bm P_C^{{(u)}}(0),k=0$
    \ForAll{$j\in \mathcal{D}^{{(d)}}$}
    \State 	BS sends $\bf B_j^{{(d)}}(0), \bf P_{C_j}^{{(d)}}(0),\bf P_{D_j}^{{(d)}}(0)$ to D2D pair $j$.
    \EndFor
    \ForAll{$j\in \mathcal{D}^{{(u)}}$}
    \State 	BS sends $\bf B_j^{{(u)}}(0), \bf P_{C_j}^{{(u)}}(0),\bf P_{D_j}^{{(u)}}(0)$ to D2D pair $j$.
    \EndFor
    \Repeat
    \State $k=k+1$
    \ForAll{$j\in \mathcal{D}^{{(d)}}$}
    \Comment{Find: $\bf P_{D_j}^{{(d)}}(k),\bf P_{C_j}^{{(d)}}(k),\bf B_j^{{(d)}}(k)$.}
    \If{PCSI mode \textbf{OR} ICSI-ERM mode}
    \State D2D pair $j$ uses the closed-form power allocation in Sec. \ref{subsec:preD}.
    \Else
    \Comment{ICSI-MRM mode}
    \State D2D pair $j$ applies \eqref{eq:FP_sol} for a single iteration.
    \EndIf
    \State D2D pair $j$ sends $\bf P_{C_j}^{{(d)}}(k)$ to the BS.
    \State 	D2D pair $j$ applies a single iteration of the PGD algorithm and sends $\bf B_j^{{(d)}}(k)$ to the BS.
    \EndFor
    \State \dots \Comment{The same for uplink}
    \State	BS projects $\bm B^{{(d)}}(k)$ and $\bm B^{{(u)}}(k)$ and sends each column to the corresponding D2D pair.
    \Until{$\bm B^{{(d)}},B^{{(u)}},\bm P_C^{{(d)}},\bm P_D^{{(d)}},\bm P_D^{{(u)}},\bm P_C^{{(u)}}$ converges}
\end{algorithmic}
\end{algorithm}
\vspace{-4mm}
\subsection{Joint Uplink and Downlink Resource Allocation}
Algorithm \ref{algo:Dec_J} below describes how this scenario can be solved in a decentralized manner. The BS initializes the power assignment vectors and the channel allocation matrices, and broadcasts them. Each D2D pair performs a step of the power allocation method suitable for the operation scenario followed by updating the vectors of the channel allocation ($\bm B^{{(d)}}_j, \bm B^{{(u)}}_j$) using a gradient step. Then, the BS assembles all the vectors of the channel allocation matrices and projects them into a feasible solution and broadcasts them to all D2D pairs. The BS and D2D pairs uses these calculated powers and channel assignments for communications. These steps are then repeated until all variables converge. Algorithm \ref{algo:Dec_J} has the same convergence and computational behaviour as Algorithm \ref{algo:Dec_I}.
\setlength{\textfloatsep}{0pt}
\begin{algorithm}[t] 
\caption{Decentralized Joint Resource Allocation} 
\label{algo:Dec_J} 
\begin{algorithmic}[1] 
    \State 	Initialize: $\bm B^{{(d)}}(0),\bm B^{{(u)}}(0),\bm P_C^{{(d)}}(0),\bm P_{D}^{{(d)}}(0), \bm P_{D}^{{(u)}}(0),$ $ \bm P_{C}^{{(u)}}(0),k=0$ 
    \ForAll{$j\in \mathcal{D}$}
    \State 	BS sends $\bf B_j^{{(d)}}(0), \bf B_j^{{(u)}}(0),\bf P_{C_j}^{{(d)}}(0),\bf P_{D_j}^{{(d)}}(0), \bf P_{D_j}^{{(u)}}(0), \bf P_{C_j}^{{(u)}}(0)$ to D2D pair $j$.
    \EndFor

    \Repeat
    \State $k=k+1$
    \ForAll{$j\in \mathcal{D}$}
    \Comment{Find: $\bf P_{D_j}^{{(d)}}(k),\bf P_{C_j}^{{(d)}}(k),\bf P_{D_j}^{{(u)}}(k),\bf P_{C_j}^{{(u)}}(k),\bf B_j^{{(d)}}(k), \bf B_j^{{(u)}}(k)$.}
    \If{PCSI mode \textbf{OR} ICSI-ERM mode}
    \State D2D pair $j$ uses the closed-form power allocation in Sec. \ref{subsec:preD}.
    \Else
    \Comment{ICSI-MRM mode}
    \State D2D pair $j$ applies \eqref{eq:FP_sol} for a single iteration.
    \EndIf
    \State D2D pair sends $P_{C_j}^{{(d)}}(k),P_{C_j}^{{(u)}}(k)$ to the BS.
    \State 	D2D pair applies a single iteration of the PGD algorithm and sends $\bf B_j^{{(d)}}(k), \bf B_j^{{(u)}}(k)$  to the BS.
    \EndFor
    \State	BS projects $\bm B^{{(d)}}(k),\bm B^{{(u)}}(k)$ and sends each column to the corresponding D2D pair.
    \Until{$\bm B^{{(d)}},\bm B^{{(u)}},\bm P_C^{{(d)}},\bm P_D^{{(d)}},\bm P_D^{{(u)}},\bm P_C^{{(u)}}$ converges}
\end{algorithmic}
\end{algorithm}
\vspace{-5mm}
\section{Simulations}
\label{sec:Sim}
We consider a simulation scenario with a single cell of radius 500 m. In this cell, CUs and D2D transmitters are located uniformly at random. The D2D receivers are located uniformly at random in a 5 m radius circle centered at their respective transmitter.
A path-loss model with exponent $\alpha=2$ is used in the calculation of all channel gains. The random channel gains are calculated by applying an exponential random distribution around an average calculated from the path-loss model. $N_C=10, N_D=10$ were used in the experiments with Monte-Carlo averages carried over $10,000$ different realizations.
\begin{figure}[!t]
\centering
\includegraphics[width=0.48\textwidth]{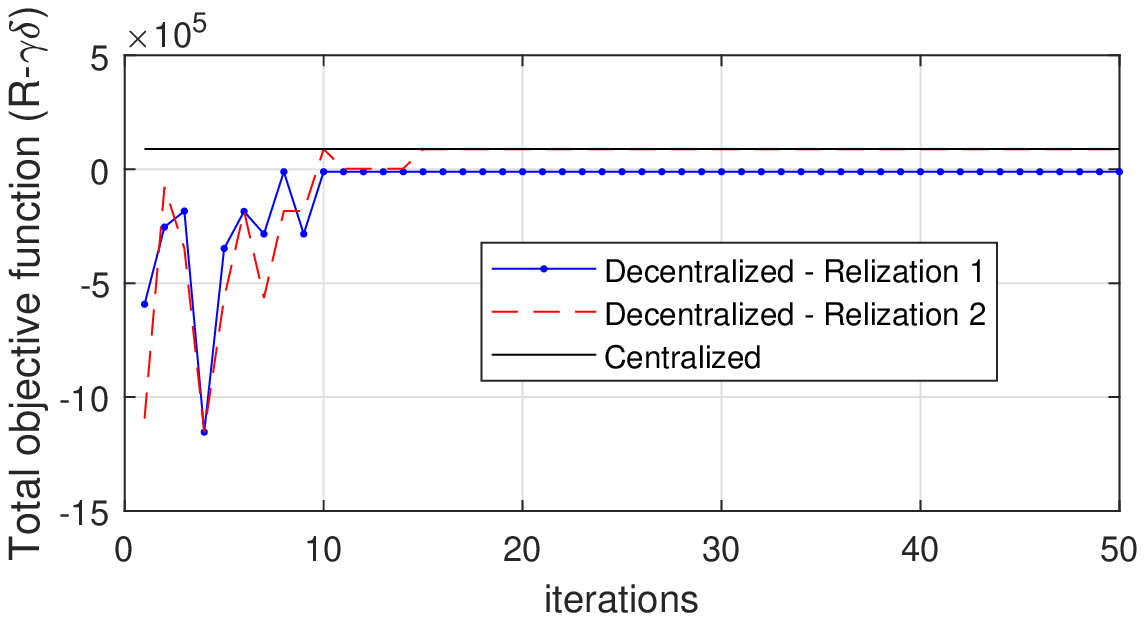}
\vspace{-5mm}
\caption{Rate of convergence}
\label{fig:Conv}
\vspace{-2mm}
\end{figure}
Fig. \ref{fig:Conv} shows the convergence results for the Decentralized Algorithm \ref{algo:Dec_J} with \gls{LCC}-\gls{MRM} compared to the Centralized Algorithm \ref{algo:Cent_J} with \gls{LCC}-\gls{MRM} of a simulation scenario with two realizations for $\gamma=50$. It shows that the decentralized algorithm converges in a relatively small number of iterations. Similar behaviour is also observed when comparing the the Decentralized Algorithm \ref{algo:Dec_I} compared to the Centralized Algorithm \ref{algo:Cent_Ind}. The obtained decentralized solutions, in general, are not identical to the centralized solution but it are very close. This is as expected because the alternating optimization for the power allocation and the binary channel allocation problem might have different solutions based on the initialization and the projection, since it is not convex.

\begin{figure}
\centering
\includegraphics[width=0.48\textwidth]{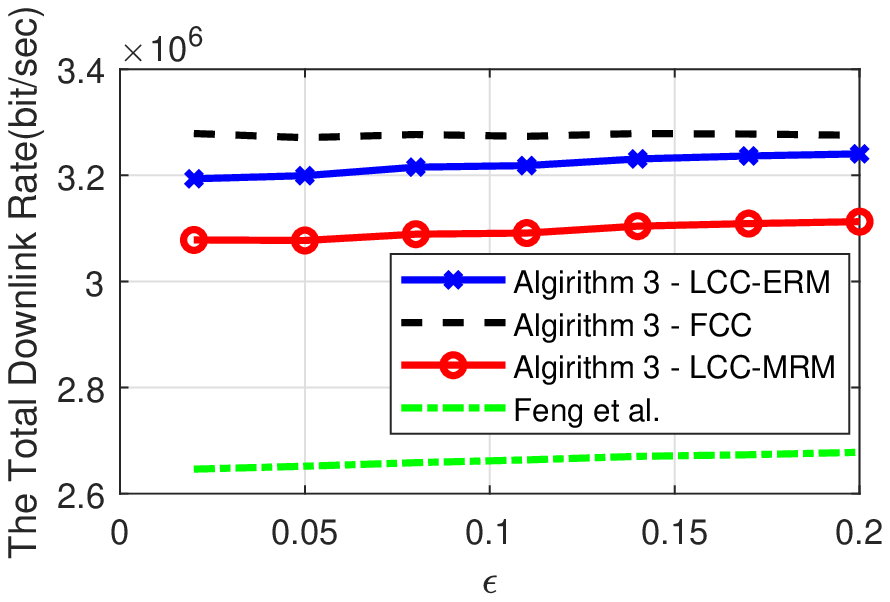}
\vspace{-3mm}
\caption{Total Rate vs. $\epsilon$}
\label{fig:M_rates}
\vspace{-7mm}
\end{figure}

\begin{figure}
\centering
\includegraphics[width=0.48\textwidth]{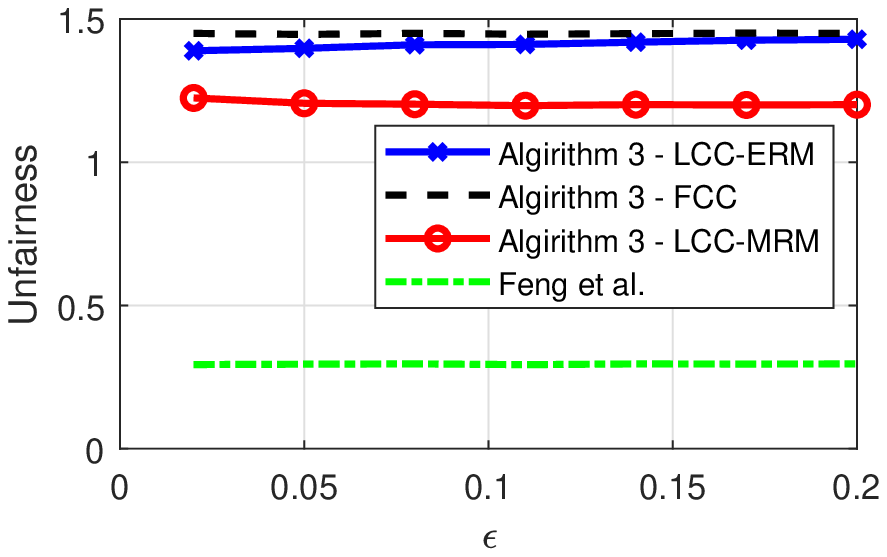}
\vspace{-5mm}
\caption{Fairness vs. $\epsilon$}
\label{fig:M_fairness}
\end{figure}

\begin{figure}
\centering
\includegraphics[width=0.48\textwidth]{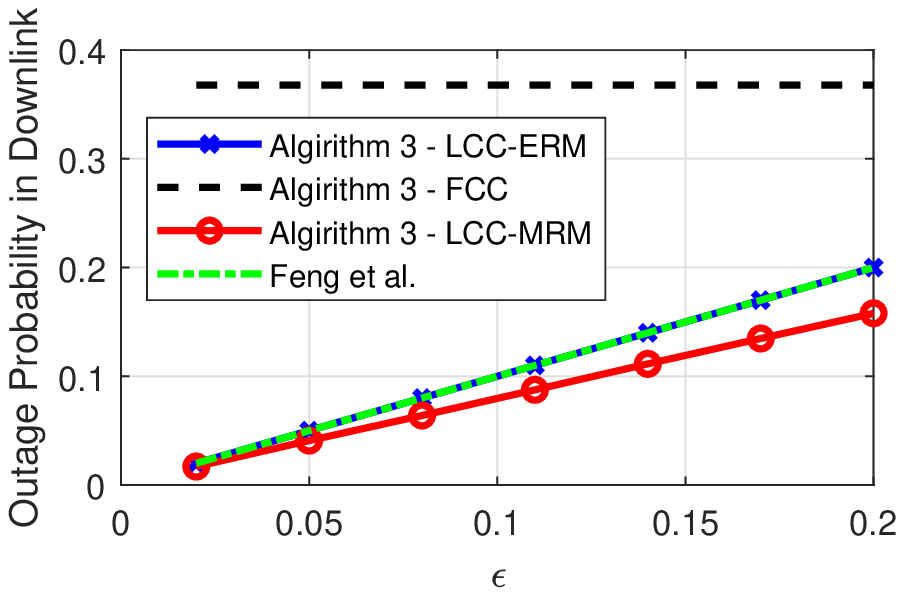}
\vspace{-3mm}
\caption{Outage Probability vs. $\epsilon$}
\label{fig:M_outage}
\vspace{-5mm}
\end{figure}

Figs. \ref{fig:M_rates}, \ref{fig:M_fairness}, and \ref{fig:M_outage} shows comparisons of Algorithm \ref{algo:Dec_I} in the cases of \gls{FCC}, \gls{LCC}-\gls{ERM} and \gls{LCC}-\gls{MRM}, compared with the previous state-of-the-art methods in \cite{feng2016qos}. Additionally, we assumed all D2D pairs will use only downlink spectrum.
The achieved rate of the \gls{FCC} case is the highest, as expected, since it ignores the probabilistic constraints and only uses the average channel gains. The cases of \gls{LCC}-\gls{ERM} and \gls{LCC}-\gls{MRM} achieve the second and third rate respectively. The method in \cite{feng2016qos} achieves the lowest rate since it does not allow assigning multiple channels to a D2D pair. The rates of all methods, except the \gls{FCC} case, grow with the allowed outage probability $\epsilon$. However, the fairness of the method in \cite{feng2016qos} is the best for the same reason (D2D pair can not access multiple channels). All cases of Algorithm \ref{algo:Dec_I} achieve relatively similar fairness, with the order of \gls{LCC}-\gls{MRM}, \gls{LCC}-\gls{ERM}, and \gls{FCC} from the second best to the forth respectively.
The achieved outage probabilities of \cite{feng2016qos} and case \gls{LCC}-\gls{ERM} are exactly equal to the allowed outage probability $\epsilon$, since the achieved optimal power assignment lies in the border of the feasibility region in both methods. Case \gls{LCC}-\gls{MRM} achieves a better outage probability than the desired $\epsilon$ with the corresponding gap increasing when $\epsilon$ increases; this can be caused by the fact that the power allocation algorithm converges to a local optima rather than the global one, and the number of feasible local optimums increases when expanding the feasibility set. The \gls{FCC} case achieves a very high outage probability, which is fixed, regardless of the value of $\epsilon$.

\begin{figure}
\centering
\includegraphics[width=0.48\textwidth]{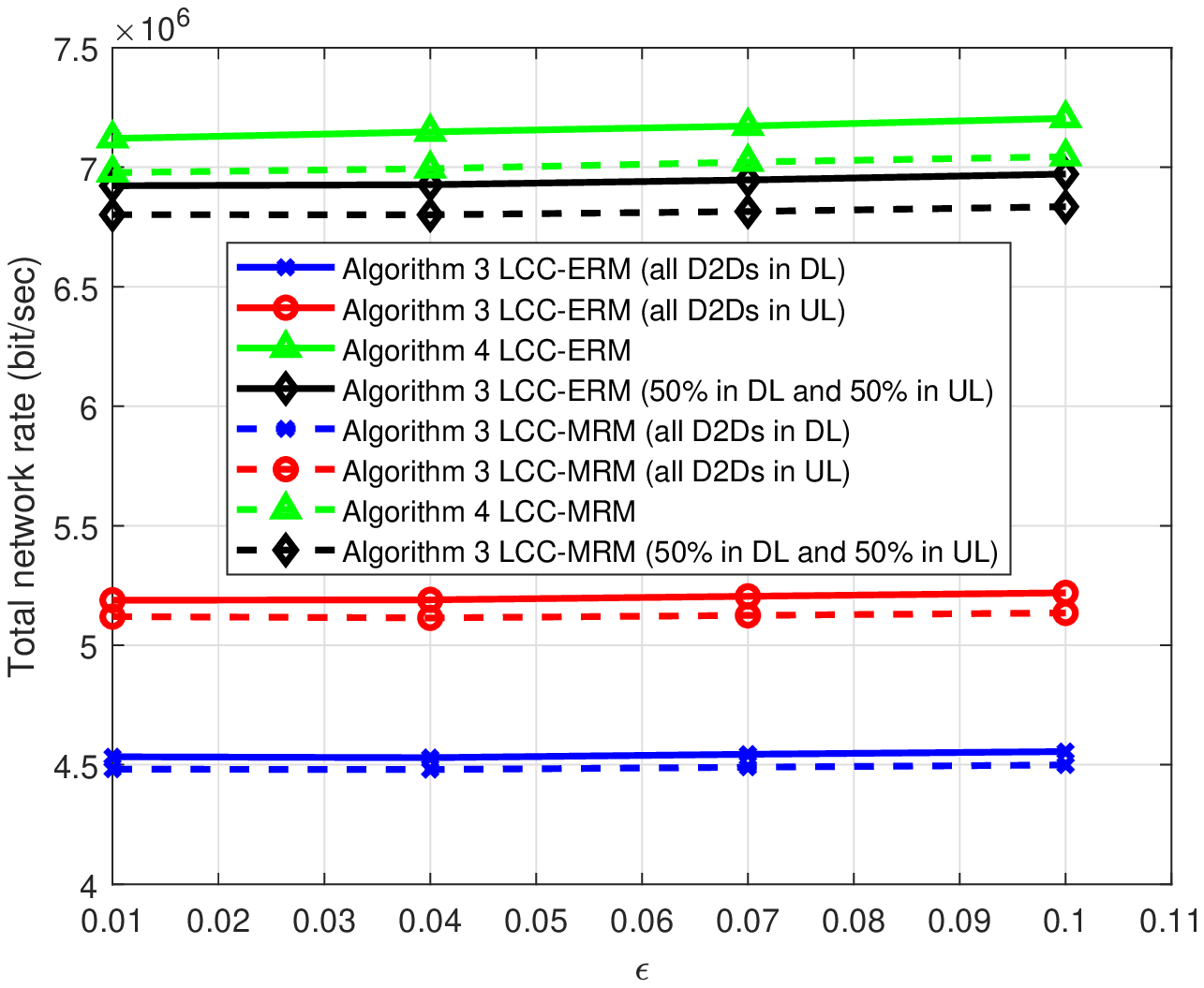}
\vspace{-5mm}
\caption{The total network rate (UL+DL) vs. $\epsilon$}
\label{fig:J_rates}
\end{figure}

Fig. \ref{fig:J_rates} shows comparisons between algorithms \ref{algo:Dec_I} and \ref{algo:Dec_J} in a ICSI-ERM case and ICSI-MRM case. 
Algorithm \ref{algo:Dec_I} is tested in the following scenarios; all D2D pairs in downlink, all D2D pairs in uplink, half D2D pairs in downlink and half in uplink. The results shows that uplink is generally better than downlink, as expected, due to the lower interference in uplink caused by the lower maximum transmitting power and the possibly longer distances between the D2D pairs and the CUs. Moreover, distributing users among both uplink and downlink achieves significantly higher data rates, with Algorithm \ref{algo:Dec_J} achieves the highest rates, since the distribution of users is also optimized.
\vspace{-5mm}
\section{Conclusion}
\label{sec:Conc}
This paper formulates a joint channel allocation and power assignment problem in underlay D2D communications. This problem aims at maximizing the total network rate while keeping the fairness among the D2D pairs. It also allows assigning multiple channels to each D2D pair. Furthermore, it assigns both downlink and uplink resources either jointly or separately. Moreover, it considers uncertainties in the CSI by including probabilistic SINR constraints to guarantee the desired outage probability.
Although this problem is a non-convex mixed-integer problem, we solve it in a computationally efficient manner by convex relaxation, quadratic transformation and alternating optimization techniques. Additionally, decentralized algorithms to solve this problem are also presented in this paper.
Numerical experiment show that our algorithms achieve substantial performance improvements as compared to the state-of-the-art.

\acom{{Future Directions}
\begin{myitemize}
\myitem Our future we will consider the possibility of having both channel assignment and channel sharing,
\myitem\cmt{MIMO} MIMO beamforming techniques in the D2D pairs under imperfect CSI,
\myitem\cmt{models} as well as incorporating user behavior models.
\end{myitemize}}
%
%
\appendices
\vspace{-5mm}
\section{Proof of Theorem 1}
\label{App:Proof1}
\acom{This alternating optimization 
generates a non-decreasing sequence $\{F_0(\mathbf{p}^{[k]})\}_{k \in N_+}$, as shown by the following inequalities:

\begin{align*}
    F_0(\mathbf{p}^{[k+1]}) & \stackrel{(a)}{=} F_1(\mathbf{p}^{[k+1]},\mathbf{z}^{[k+1]}) 
     \stackrel{(b)}{\geq} F_1(\mathbf{p}^{[k+1]},\mathbf{z}^{[k]}) 
     \stackrel{(c)}{=} F_2(\mathbf{p}^{[k+1]},\mathbf{z}^{[k]},\tilde{\mathbf{y}}^{[k]}) \\
    & \stackrel{(d)}{\geq} F_2(\mathbf{p}^{[k+1]},\mathbf{z}^{[k]},\mathbf{y}^{[k]}) 
     \stackrel{(e)}{\geq} F_2(\mathbf{p}^{[k]},\mathbf{z}^{[k]},\mathbf{y}^{[k]}) 
     \stackrel{(f)}{=} F_1(\mathbf{p}^{[k]},\mathbf{z}^{[k]}) 
     \stackrel{(g)}{=} F_0(\mathbf{p}^{[k]})
\end{align*}
Here, equalities $(a)$ and $(g)$ follow from the fact that $F_1(\mathbf{p},\mathbf{z}^*) = F_0(\mathbf{p})$, and equalities $(c)$ and $(f)$ follow from $F_2(\mathbf{p},\mathbf{z},\mathbf{y}^*) = F_1(\mathbf{p},\mathbf{z})$; inequality $(b)$, $(d)$, and $(e)$ follow from the fact that $F_1(\mathbf{p},\mathbf{z}^*) \geq F_1(\mathbf{p},\mathbf{z})$ \myacom{as can be seen in \eqref{eq:d4}} and $F_2(\mathbf{p},\mathbf{z},\mathbf{y}^*) \geq F_2(\mathbf{p},\mathbf{z},\mathbf{y})$, where $(\bm.)^*$ denotes the optimal solution for the respective sub-problem.}
\acom{Next, given $\{F_0(\mathbf{p}^{(k)})\}_{k \in N_+}$ is a non-decreasing sequence, a sufficient condition for $\{\mathbf{p}^k\}_{k \in N_+}$ to converge to a stationary point is that the set $\{\mathbf{p} \in \mathcal{P} : F_0(\mathbf{p}) \geq F_0(\mathbf{p}^0)\}$ is a compact set \cite{scutari2018parallel}. 
Furthermore, notice that the value of $F_0(\mathbf{p})$ is bounded above, thus,  the set $\{\mathbf{p} \in \mathcal{P} : F_0(\mathbf{p}) \geq F_0(\mathbf{p}^0)\}$ is compact, and hence the sequence $\{\mathbf{p}^k\}_{k \in N_+}$ converges to a stationary point.}
First, let us define an equivalent problem to \eqref{eq:d5} as follows:
{\small
\begin{subequations}
\begin{align}
\label{eq:d5e}
\maximize_{\bm p,\bm{z},\bm y}\quad &F_2(\bm p,\bm{z},\bm y)
\\\nonumber
\text{subject to } \quad &\eqref{eq:d2rmaxp}, ~\eqref{eq:d2rSINR},\\
\label{eq:sol_zs_E}
&z_1=\dfrac{P_{C_{ij}}g_{C_i}}{N_0+P_{D_{ji}}\text{F}^{-1}_{\tilde{h}_{D_{j,i}}}(1-\epsilon)},~ \quad z_2=\dfrac{P_{D_{ji}}g_{D_j}}{N_0+P_{C_{ij}}h_{C_{j}}},\\
\label{eq:sol_ys_E}
&y_1=\dfrac{\sqrt{(1+z_1)P_{C_{ij}}g_{C_i}}}{P_{C_{ij}}g_{C_i}+N_0+P_{D_{ji}}\text{F}^{-1}_{\tilde{h}_{D_{j,i}}}(1-\epsilon)},~
 y_2 =\dfrac{\sqrt{(1+z_2)P_{D_{ji}}g_{D_j}}}{P_{D_{ji}}g_{D_j}+N_0+P_{C_{ij}}h_{C_{j}}},
\end{align}
\end{subequations}}
where the additional constraints \eqref{eq:sol_zs_E} and \eqref{eq:sol_ys_E} are obtained from the solutions of \eqref{eq:d5} in \eqref{eq:sol_zs} and \eqref{eq:sol_ys} respectively. Since the solution of \eqref{eq:d5} lies in the feasible set of \eqref{eq:d5e}, both problems are equivalent.

Next, we prove that the limit point $\bar{\mathbf{p}}$ of $\{\mathbf{p}^{[k]}\}_{k \in \mathbb N_+}$ is a stationary point of \eqref{eq:d5e}. It is shown in Theorem 2.8 in \cite{xu2013block} that, for any bounded continuous function that (i) is \emph{locally Lipschitz smooth and strongly convex} for each block in the feasibility set, (ii) has a Nash point, and (iii) satisfies the Kurdyka Lojasiewicz (KL) property in a neighborhood around a stationary point, the sequence generated by an alternation optimization algorithm with a fixed update scheme initialized in that neighborhood will converge to that stationary point. We will next show that $-F_2$ satisfies (i), (ii), and (iii).

A Nash point $\bar{\bold x}$ for a function $f(\bold x)$ is defined as a block-wise minimizer where $f(\bar{\bold x}_1,\dots,\bar{\bold x}_i,$ $\dots,\bar{\bold x}_s)$ $\leq f(\bar{\bold x}_1,\dots,\bold{\bold x}_i,\dots,\bar{\bold x}_s)~\forall i$ and $\bold x=(\bold x_1,\bold x_2,\dots,\bold x_s)$ \cite{xu2013block}.
Since $F_2$ is continuous and the feasible set of \eqref{eq:d5e} is compact, $F_2$ attains a locally optimal point as stated by Weierstrass’ Theorem described in (A.2.7) in \cite{bertsekas2009convex}. Thus, the function $F_2$ has a Nash point.
The function $-F_2(\mathbf{p},\mathbf{z},\tilde{\mathbf{y}})$ can be shown to be strongly convex and Lipschitz smooth in each variable separately in the bounded feasible set of \eqref{eq:d5e} ({\small$\partial^2 F_2/\partial z_1^2=-(1+z_1)^{-2}-y_1\sqrt{P_{C_{ij}}g_{C_i}}(1+z_1)^{-3/2}/2$, $\partial^2 F_2/\partial z_2^2=-(1+z_2)^{-2}-y_2\sqrt{P_{D_{ji}}g_{D_i}}(1+z_2)^{-3/2}/2$, $\partial^2 F_2/\partial y_1^2=-2(P_{C_{ij}}g_{C_i}+N_0+P_{D_{ji}}\text{F}^{-1}_{\tilde{h}_{D_{j,i}}}(1-\epsilon))$, $\partial^2 F_2/\partial y_2^2=-2(P_{D_{ji}}g_{D_j}+N_0+P_{C_{ij}}h_{C_{j}})$, $\partial^2 F_2/\partial P_{C_{ij}}^2=-y_1\sqrt{(1+z_1)g_{C_i}}(P_{C_{ij}})^{-3/2}/2$, $\partial^2 F_2/\partial P_{D_{ji}}^2=-y_2\sqrt{(1+z_2)g_{D_j}}(P_{D_{ji}})^{-3/2}/2$}). 

To see that (iii) holds, we use the following definition of the Kurdyka Lojasiewicz (KL) property \cite{xu2013block}:
\vspace{-4mm}
\theoremstyle{definition}
\begin{definition}{Kurdyka Lojasiewicz (KL) property:}
A function $f(\bold x)$ satisfies the KL property at point $\bar{\bold x} \in \text{dom}(\partial f)$ if $\eta=\dfrac{|f(\bold x)-f(\bar{\bold x})|^\theta}{dist(\bold 0,\partial f(\bold x))}$ is bounded for $0\leq \theta<1$ $\forall \bold x$ in some neighborhood $U$ of $\bar{\bold x}$.
\end{definition}
\vspace{-5mm}
We then introduce a lemma as follows:
\vspace{-5mm}
\begin{lemma}
The function $F_2(\mathbf{p},\mathbf{z},\mathbf{y})$ satisfies the KL property at any point $\mathbf{p} \in \mathcal{P}$, $\mathbf{z} \in \R^2_+$ and $\mathbf{y} \in \R^2_+$, for some $\theta \in [1/2,1)$.
\end{lemma} \vspace{-5mm}
\textit{Proof}: see Appendix \ref{App:Proof2}\\
Since $F_2$ is analytic everywhere, it is also analytic around a stationary point $[\bar{\bold p}, ~\bar{\bold z}, ~\bar{\bold y}]^T$ and, consequently, satisfies (iii).
Thus the alternation sequence in \eqref{eq:FP_sol} initialized at any feasible point $\bold p^0$ will converge to the nearest stationary point of $F_2$, since $\bold p^0$ is in the neighborhood of the nearest stationary point $\bar{\bold p}$. 
Moreover, any stationary point of $F_2$ is a stationary point of $F_0$ \cite{shen2018fractional,shen2018fractional2}. Thus the sequence $\{\mathbf{p}^{[k]}\}_{k \in N_+}$ converges to a stationary point of $F_0$.
%

Next, we prove that $\vert \mathbf{p}^{[k]}-\bar{\mathbf{p}}\vert\leq C k^{-(1-\theta)/(2\theta-1)}$ for some $C>0$.
It is shown in theorem 2.9 in \cite{xu2013block}, for a function $f(\bold x)$ that satisfies Theorem 2.8 in \cite{xu2013block} and the KL property for some $\theta \in (1/2,1)$, the update sequence ${\bold x^{[k]}}$ converges to a stationary point $\bar{\bold x}$ as $\vert \bold x^{[k]}-\bar{ \bold x}\vert\leq C k^{-(1-\theta)/(2\theta-1)}$ with a certain $C>0$. Since $ F_2(\mathbf{p},\mathbf{z},\mathbf{y})$ satisfy the KL property for some $\theta \in (1/2,1)$, the update sequence $\{\mathbf{p}^k\}_{k \in N_+}$ converges as $\vert \bold{p}^{[k]}-\bar{ \bold p}\vert\leq C k^{-(1-\theta)/(2\theta-1)}$ to a stationary point $\bar{\mathbf{p}}$. 
\vspace{-5mm}
\section{Proof of Lemma 1}
\label{App:Proof2}
%
It can be shown that any real analytic function satisfies the KL property for some $\theta \in [1/2,1)$ \cite[sec. 2.2]{xu2013block}.
Next, we need to show that $ F_2(\mathbf{p},\mathbf{z},\mathbf{y})$ is a real analytic function. A function $f(\bold x)$ is a real analytic function if it is infinitely differentiable and its Taylor series around a point $\bold x_0$ converges to $f(\bold x)$ for $\bold x$ in some neighbourhood of $\bold x_0$ \cite{absil2005convergence}. To simplify our problem, we first need to consider the following properties of real analytic functions \cite{absil2005convergence}:
    (i) The sum and product of real analytic functions is a real analytic function.
    (ii) Any polynomial is a real analytic function.
    (iii) The composition of real analytic functions is a real analytic function.
Exploiting, the first property, it suffices to show that all individual terms in the expression of $F_2(\mathbf{p},\mathbf{z},\mathbf{y})$ in \eqref{eq:d5} are real analytic. 

We first show that 
$\log(\bm.)$ is real analytic on a positive real argument, i.e., 
$x \in (0, \infty)$.
This can be formally proved by showing that the remainder of the order-n Taylor series expansion of $\log(x)$ centered around a point $c$ goes to zero as $n$ goes to infinity. 
The Taylor series expansion of $\log(x)$ centered at $c > 0$, can be expressed as:
    $\sum^{\infty}_{n=0} \frac{(-1)^n(n-1)!}{n!c^n}(x-c)^n.$
Our objective is to show that above expansion converges to $\log(x)~\forall ~x\in (c/2,3c/2)$. The Lagrange reminder of the Taylor's series expansion of function $f(x)$ can be expressed as:
  $ R_n(x) = \frac{f^{(n+1)}(\zeta)}{(n+1)!} (x-c)^{n+1}$, 
where, $f^{(n+1)}(.)$ is the $(n+1)$-th derivative of $f$. Substituting $f(\zeta) = \log(\zeta)$, we have
   $R_n(x) = \frac{(-1)^n n!}{(n+1)!\zeta^{n+1}} (x-c)^{n+1}$, 
where, $x, \zeta \in (c/2,3c/2)$. Simplifying further,
 $  |R_n(x)| = \frac{1}{(n+1)!}\frac{|x-c|^{n+1}}{|\zeta|^{n+1}} \leq \frac{1}{n+1} $. 
Thus, $\lim_{n \rightarrow \infty}$, $|R_n(x)| \rightarrow 0$. Hence, Taylor's series expansion of $\log(x)$ centered at $c$ converges to $\log(x)$ on $(c/2,3c/2)$. Further, if $c \rightarrow \infty$, $\log(x)$ is real analytic for $x \in (0,\infty)$. Thus, $\log(1+z_1)$ and $\log(1+z_2)$ are real analytic functions for $z_1, z_2 > 0$. \\
Next, we consider the following terms of $ F_2(\mathbf{p},\mathbf{z},\mathbf{y})$: $z_1; z_2; y_1^2(p_Bg_D + N_0 + P_Dh_D); y_2^2(p_Dg_D + N_0 + P_Bh_B)$. It can be noted that all these terms are positive polynomials. Thus, by the second property, all of these terms are real analytic functions.
Finally, for the terms $2y_1\sqrt{(1+z_1)p_Bg_B}$ and $2y_2\sqrt{(1+z_1)p_Bg_B}$, we exploit the first and the third properties. Note that $y_1,y_2$, $(1+z_1)p_Bg_B$ and $(1+z_1)p_Bg_B$ are positive polynomials; hence, they are real analytic functions. Thus, we just need to show that the square root is a real analytic function. Let $f(x): \R \rightarrow \R$ be a real analytic function. Then,
    $\sqrt{f(x)} = e^{\log\left(\sqrt{f(x)}\right)} = e^{\frac{\log(f(x))}{2}}$. 
Since the composition of real analytic functions is real analytic; given that $e^{(.)}$ is real analytic and $\log(f(x))$ is real analytic for $f(x) > 0$; then, we can conclude that $\sqrt{(f(x))}$ is real analytic for $f(x) > 0$. 
\vspace{-3mm}
\if\editmode1 
\onecolumn
\printbibliography
\else
{\small
\bibliography{\bibfilenames}}
\fi
\end{document}

%% file: include.tex
\usepackage{fixltx2e}

\usepackage{graphicx}

\usepackage{subcaption}              

\usepackage[utf8]{inputenc}

\usepackage{amsfonts}
\usepackage{amsmath}
\usepackage{mathtools}

\usepackage{amssymb}

\usepackage{bm}

\usepackage{color,verbatim}
\usepackage{multirow}
\usepackage{accents}

\usepackage{theoremref}

\usepackage{url}

\newcounter{rulecounter}
\newcommand{\resetrule}{ \setcounter{rulecounter}{0}}
\resetrule

\newsavebox{\selvestebox}
\newenvironment{colbox}[1]
  {\newcommand\colboxcolor{#1}%
   \begin{lrbox}{\selvestebox}%
   \begin{minipage}{\dimexpr\columnwidth-2\fboxsep\relax}}
  {\end{minipage}\end{lrbox}%
   \begin{center}
   \colorbox{\colboxcolor}{\usebox{\selvestebox}}
   \end{center}}

\definecolor{orange}{rgb}{1,0.8,0}
\definecolor{gray}{rgb}{.9,0.9,0.9}
\definecolor{darkgray}{rgb}{.3,0.3,0.3}
\definecolor{darkblue}{rgb}{.1,0.0,0.3}
\definecolor{lightblue}{rgb}{0.7,0.7,1}
\definecolor{lightred}{rgb}{1,0.7,.7}
\definecolor{purple}{RGB}{204,153,255}
\definecolor{lightgray}{rgb}{.95,0.95,0.95}
\definecolor{lightgreen}{rgb}{0.3,0.5,0.3}
\definecolor{darkgreen}{rgb}{0.05,0.3,0.05}

\newcommand{\ra}{$\rightarrow$~}

\newcommand{\brackets}[1]{\left\{#1\right\}}

\newcommand{\var}[1]{\mathop{\textrm{Var}}\brackets{#1} }

\newcommand{\maximize}{\mathop{\text{maximize}}}

\DeclareMathOperator*{\argmax}{arg\,max}

\newtheorem{myproposition}{Proposition}
\newtheorem{myremark}{Remark}
\newtheorem{myproblemstatement}{Problem Statement}
\newtheorem{mylemma}{Lemma}
\newtheorem{mytheorem}{Theorem}
\newtheorem{mydefinition}{Definition}
\newtheorem{mycorollary}{Corollary}